\documentclass{llncs}

\usepackage{amsmath}
\usepackage{amsfonts}
\usepackage{amssymb}
\usepackage{cancel}
\usepackage{caption}
\usepackage{cite}
\usepackage{esdiff}
\usepackage{floatrow}
\usepackage{graphicx}
\usepackage{hyperref}
\usepackage[utf8]{inputenc}
\usepackage{libertine}
\usepackage{listings}
\usepackage{lscape}
\usepackage{mathtools}
\usepackage[version=3]{mhchem}
\usepackage{multicol}
\usepackage{physics}
\usepackage{quotchap}
\usepackage{ragged2e}
\usepackage{siunitx}
\usepackage{subcaption}
\usepackage{tensor}
\usepackage{todonotes}
\usepackage{wasysym}
\usepackage{wrapfig}
\usepackage{xcolor}
\usepackage{xparse}

\newfloatcommand{capbtabbox}{table}[][\FBwidth]

\title{Brain Structural Saliency Over The Ages}

\author{Daniel Taylor \inst{1} \and Jonathan Shock \inst{1, 2, 3} \and Deshendran Moodley \inst{1} \and Jonathan Ipser \inst{1} \and Matthias Treder \inst{4}}

\institute{University of Cape Town, South Africa \and The National Institute for Theoretical and Computational Sciences, South Africa \and Institut National de la Recherche Scientifique, Montreal, Canada \and Cardiff University, Wales}

\begin{document}
    
    %\mainmatter

    \maketitle
    
    \begin{abstract}
        Brain Age (BA) estimation via Deep Learning has become a strong and reliable bio-marker for brain health, but the black-box nature of Neural Networks does not easily allow insight into the features of brain ageing. We trained a ResNet model as a BA regressor on T1 structural MRI volumes from a small cross-sectional cohort of $524$ individuals. Using Layer-wise Relevance Propagation (LRP) and DeepLIFT saliency mapping techniques, we analysed the trained model to determine the most relevant structures for brain ageing for the network, and compare these between the saliency mapping techniques. We show the change in attribution of relevance to different brain regions through the course of ageing. A tripartite pattern of relevance attribution to brain regions emerges. Some regions increase in relevance with age (e.g. the right Transverse Temporal Gyrus); some decrease in relevance with age (e.g. the right Fourth Ventricle); and others are consistently relevant across ages. We also examine the effect of the Brain Age Gap (BAG) on the distribution of relevance within the brain volume. It is hoped that these findings will provide clinically relevant region-wise trajectories for normal brain ageing, and a baseline against which to compare brain ageing trajectories.
    \end{abstract}
    
    \section{Introduction}
    
        Brain Age (BA) prediction by Deep Learning (DL) methods using structural Magnetic Resonance Imaging (MRI) data has shown to be an accurate and reliable bio-marker for brain health \cite{Cole2017}. The Brain Age Gap (BAG) of an individual -- the difference between the predicted BA and the chronological age -- is an increasingly popular and predictive bio-marker for brain health, and has been shown to predict accelerated or slowed brain ageing \cite{Smith2019, Dinsdale2021, Peng2021}.
        
        DL frameworks have shown great efficacy in the BA regression task \cite{Cole2018, Jonsson2019, Kossaifi2020, Peng2021}. An advantage of DL models is that they are able to analyse whole-brain structural images with minimal pre-processing. Other summary statistics such as measures of cortical thickness, volume and surface area \cite{Zhao2019} treat the brain as being modular, and may therefore lack details contained in minimally-processed MRI volumes (which represent the entire structure with all its integrated substructures). Convolutional Neural Networks (CNNs) have been shown to provide extremely small Mean Absolute Error (MAE) between true age and BA. The state-of-the-art in BA regression at the time of writing is around $2.5$y MAE \cite{Kossaifi2020, Dinsdale2021, Peng2021}. Such accuracies have thus far only been achieved on large datasets, and some of these datasets are limited in their age ranges, such as the UK BIOBANK \cite{Sudlow2015} (45y-80y).
        
        The black-box nature of Neural Networks makes it difficult to attribute BAG to specific brain regions. Many methods of Explainable Artificial Intelligence (XAI) have aimed at alleviating this issue. These attempt to explain why the models make their predictions. A popular post-hoc method of reasoning for DL models is saliency mapping. This is a group of methods by which areas of input are assigned relevance scores proportional to their saliency to the model decision. We compare the results of two widely used saliency mapping methods. We explore Layer-wise Relevance Propagation (LRP) \cite{Bach2015} and DeepLIFT \cite{Shrikumar2017} as saliency mapping methods for the BA task. These methods were chosen due to their past performance in the literature in DL-based brain imaging tasks \cite{Gupta2019, Grigorescu2019, Eitel2019, Levakov2020, Hofmann2021}. In this work, we use the terms `salience' and `relevance' interchangeably.
        
        Many factors can contribute to accelerated BA (positive BAG), such as the presence of neurodegenerative disease, type 2 diabetes or HIV, and past physical activity \cite{Cole2017}. If we wish to create a model that accurately predicts BA, and therefore accurately assesses BAG, we must ensure that the model accurately captures a path of `normal' brain ageing. It must thus be trained on data in which the subjects do not have any underlying pathologies that can affect BA independent of actual age (to the extent that such pathologies are able to be detected). Even in a `healthy' cohort, however, there will be non-negligible deviation from an average ageing trajectory. Smith et al. \cite{Smith2019} note that some meaningful BAG should exist between the predicted and chronological age of an individual, as long as the model does not badly over-fit.
        
        In this work we aim to shed light on the contributions of specific brain regions to BA and BAG through the course of ageing. Specifically we apply saliency mapping techniques to a BA regression model and compare the results to known characteristics of brain ageing. We then analyse the differences between saliency mapping techniques' distributions of relevance throughout the brain volume. We examine the link between region-specific saliency and accelerated brain ageing both in older and younger individuals. Finally, we create region-specific trajectories of BA saliency across age groups from a population study.
        
        The primary contribution of this research is the development of methods to determine trajectories of BA saliency over the course of age; allowing us not only to determine the saliency of the region towards age and the change of this through time, but also to create baseline saliency trajectories against which to compare and assess BA of individuals at a region-specific level. Another contribution of the work is the analysis of regional saliency in individuals with accelerated BA (large BAG), which allows us to determine key contributing regions to accelerated BA in younger and older adults. Our final contribution is the comparison of different saliency mapping techniques, allowing us not only to compare the similarities and differences between them, but also to shed light on different aspects of BA through these differences.
        
        %We explore the trajectories of relevance assignment across ages, comparing multiple saliency mapping techniques. It is hoped that this work will provide a clinically relevant baseline of salience trajectories for brain regions with respect to BA.
        
        From our findings in the current literature, we expected that regions that are deemed highly relevant in general towards BA would increase in their relevance with age. Areas that are generally less salient would then necessarily decrease in assigned relevance.
        
    \section{Related Work}
	    
	    Levakov et al. \cite{Levakov2020} used SmoothGRAD \cite{Smilkov2017} to create population-level saliency maps for a BA regression model. The authors examined which brain regions corresponded with the highest attribution of saliency by thresholding the top-$1\%$ of voxels in saliency maps and examining clusters with $>100$ voxels. Most salience was attributed to the cisterns and ventricles.
	    
	    Hofmann et al. \cite{Hofmann2021} used LRP to create saliency maps for two multi-ensemble BA regression models. They note that the SmoothGRAD method employed by \cite{Levakov2020} is not directional, while LRP highlights areas of input that both agree with and contradict the output. They utilised the LRP$_\text{CMP}$ saliency method with $\alpha=1$ (see Appendix (\ref{app:saliency_maps})). The authors verified the method on a simulated brain ageing model and found that in the regression task, regions of higher relevance argue for greater age, while regions of lower relevance argue for lower age. After verifying the method, the authors created the LRP saliency maps for the real BA regression task. Like \cite{Levakov2020} they found that relevance was greatest in and around the ventricles. They also found a significant attribution to the grey-matter-dense regions at the cortical surface.The authors also contrasted the saliency maps of individuals in a younger and an older cohort to determine the difference in attribution of relevance.
	    
	    There have been no previous attempts in the literature to compare the results of different saliency mapping techniques on the BA regression task. While \cite{Hofmann2021} compared a younger and older cohort in their study, no previous work has focused on the change in relevance attributions to brain regions as a function of subject age. The authors also only examined statistical significance toward BAG in an older cohort, and not age-related contributors to BAG. Furthermore, the concern raised by Geirhos et al. \cite{Geirhos2019} and Sixt et al. \cite{Sixt2019} about modified backpropagation algorithms like LRP has not been addressed. Their concern is that such methods attempt to recreate the input to the model, as opposed to focusing solely on areas of relevance.
	
	\section{Methods}
	    
	    \subsection{BA Regression Model}
	    
    	    The model used in the BA regression task was a ResNet \cite{He2016} with filter number sequence $\left[32, 32, 64, 64, 128\right]$ in the main branches of the 5 residual blocks. This configuration of residual blocks is borrowed from \cite{Jonsson2019}. We used a softplus activation function for all the nonlinearities in the model, as recommended by Dombrowski et al. \cite{Dombrowski2019}, for the robustness of the saliency maps.
    	    
    	    We used the cross-sectional Cam-CAN T1-weighted MRI dataset \cite{Shafto2014, Taylor2017} for our experiments. The volumes are from 656 healthy individuals ranged 18-89y, with $49.4\%$ or $324$ male participants.
    	    
    	    We trained our ResNet model on a random $80\%$ training split of the dataset (524 subjects). 80 of these were randomly held out for validation ($15\%$ validation split from the beginning of training). The remaining $20\%$ of the dataset (132 subjects) was used in testing. We used grid-search hyper-parameter tuning to find the best optimiser for the model (Adam \cite{Kingma2014} vs RMSProp), as well as the best loss function (Mean Squared Error vs Mean Absolute Error) and the best starting learning rate ($5\times10^{-4}$ vs $10^{-3}$ vs $5\times10^{-3}$). Taking the average of three trials for each hyper-parameter configuration, it was found that the best configuration was using the Mean Squared Error (MSE) loss function with the Adam optimiser and an initial learning rate of $5\times10^{-3}$. For a set of $n$ predictions $\left[f(x_i)\right]$ and the corresponding set of $n$ true ages $[y_i]$, the MSE is defined as $\text{MSE}\left(\left[f(x_i)\right], [y_i]\right) = \frac{\sum_i\left(f(x_i) - y_i\right)^2}{n}$. We did not perform hyper-parameter tuning on any other hyper-parameters as the rest of the model configuration was based off of the model of \cite{Jonsson2019}. The learning rate was halved every 20 epochs using a learning rate schedule. We also utilised an early stopping callback in training such that if there was no validation improvement for 20 epochs, the training would cease. Due to the large size of the data ($233\times189\times197$ voxels per scan at $1\text{mm}^3$ resolution), the batch size was limited to 4. The metrics used to evaluate the model were the MAE, MSE and Mean Average Percentage Error (MAPE). Appendix (\ref{app:preprocessing}) details the pre-processing of the data.
        
        \subsection{Saliency Mapping}
	    
	        We performed saliency mapping to analyse the trained model using LRP \cite{Bach2015, Montavon2017} and DeepLIFT \cite{Shrikumar2016, Shrikumar2017}. We used three variations of LRP$_\text{CMP}$ (as this method is considered to be best-practice \cite{Kohlbrenner2019}), with the parameters $\alpha\in\left\{1, 2, 3\right\}$ (the same values implemented by Grigorescu et al. \cite{Grigorescu2019}). We refer to these by the shorthand LRP$_1$, LRP$_2$ and LRP$_3$. In all three cases, we used the \textit{Epsilon Rule} for dense layers and the $z^\mathcal{B}$\textit{-Rule} for the input layer \cite{Montavon2017}. We used two variations of the DeepLIFT method. In the first case ($_\text{bg}$), we used a reference input of all zeros (the MRI background). In the second case ($_\text{comp}$) we used as reference a composite MRI volume, which was the aggregate of all the subjects' volumes from the test set. For the nonlinear softplus activation layers, we used the Reveal-Cancel Rule \cite{Shrikumar2017}, which was found to produce more consistent and interpretable saliency maps than the Rescale Rule. Both DeepLIFT methods allow for positive and negative relevance assignment.
            
            For all methods, we used permutation-based t-tests to determine statistically significant differences from the input volumes. This is in order to address the concern of \cite{Geirhos2019} that the LRP methods simply recreate the input.
            
            \subsubsection{Regional Relevance Attribution Between Methods}
            
                To compare the saliency mapping methods, we examined the attribution of top-$1\%$ relevance (T1R) to different brain areas across the test set, per method. T1R refers to the voxels with the first percentile of activations in the brain volume for a saliency map. This is similar to the methodology used in \cite{Levakov2020}. We count the number of such voxels per region, and normalise by the region volume to get the proportion of the region assigned T1R. We assessed the similarities and differences in regional relevance attribution between the methods, to determine how each of the methods explains BA.
                
                To determine which brain regions saliency voxels were attributed to, we used the CerebrA 2009c atlas \cite{Collins1999, Fonov2009} and a corresponding standard MNI brain volume \cite{Manera2020} to which all of the dataset volumes were aligned before training. This allowed us to map regions directly onto each individual's brain volume (and thus the corresponding saliency map) to determine the proportion of T1R per region. It is by this method that we are able to compare the regional attributions of T1R among participants.
                
                From the findings of \cite{Hofmann2021}, we know that regions of high saliency argue for higher age and those of lower saliency argue for lower age in the context of BA regression, and so the directionality of relevance is important. We are most interested in regions which contribute to accelerated brain ageing, and so we consider only positive relevance in T1R (see the discussion of directionality by \cite{Hofmann2021}).
                
            \subsubsection{Regional Relevance Attribution Based on BAG}
                
                We used the $\delta_2$ definition of Smith et al. \cite{Smith2019}, which uses an orthogonality matrix multiplication on the traditional chronological-predicted BA difference to remove dependence of the BAG on chronological age.
                
                To examine the effect of large BAG on the distribution of relevance in the brain, we thresholded BAG in the test set and compared the T1R distributions of those lying above and below the threshold $\delta^*$. This has not been done before in the literature, and so we made the simple choice of a threshold for BAG of $\delta^*=\sigma$, where $\sigma$ is the standard deviation of the BAG values for the test set.
                We compared the distribution of T1R within the brains of those with BAG $\delta$ lying above the threshold ($\delta \geq \delta^*$) to those lying within the threshold ($\left|\delta\right| < \delta^*$).
                To calculate BAG, we used the age-orthogonal value $\delta_2$ of \cite{Smith2019}. We compare the T1R distributions of three groups for each method: $\delta_2 \geq \delta^*$ for younger individuals ($<50$y), $\left|\delta_2\right| < \delta^*$, and $\delta_2 \geq \delta^*$ for older individuals ($>50$y).
                
            \subsubsection{Regional Relevance Attribution Across Age Brackets}
                
                To assess the change in the distribution of relevance as a function of age, we examined how the T1R assignment for each method changes between equally spaced age brackets. We did this for each region individually, by calculating the average proportion of T1R assigned to the region across the test set members who lie within each age bracket.
                
                Of greatest interest were highly-relevant regions whose relevance changes most or least across ages. To quantify the greatest change, we examined those with the highest standard deviation (SD) in the proportion that is assigned T1R. To quantify the least change, we examined those with lowest coefficient of variation (CoV -- standard deviation normalised by mean).
                
                We did not focus on those regions with lowest SD since those tend to be the regions with lowest mean relevance attribution. Similarly we did not focus on those with highest CoV, since these also tend to be regions with low mean attributions, where a small change in relevance attribution can lead to a very large CoV. Our aim was to focus on the changes in relevance for highly-relevant structures. More is given on this choice of metrics in Appendix (\ref{app:dba}).
	
	\section{Results}
	    
	    \subsection{BA Regression Model}
	        
	        The training was stopped by the early stopping callback at the end of the 143$^\text{rd}$ epoch, at which pint it had reached a test MAE of $6.55$y, with an MSE of $72.55\text{y}^2$ and an MAPE of $13.53$. The performance on the evaluation metrics is given in Table (\ref{tab:model_metrics}), and the regression plot is shown in Figure (\ref{fig:test_regression}); both in Appendix (\ref{app:regression_model}).
	    
	    \subsection{Saliency Mapping}
    	    
\begin{figure}[htbp]
                \centering
                \begin{subfigure}[b]{0.45\textwidth}
    	            \centering
    	            \includegraphics[width=\linewidth]{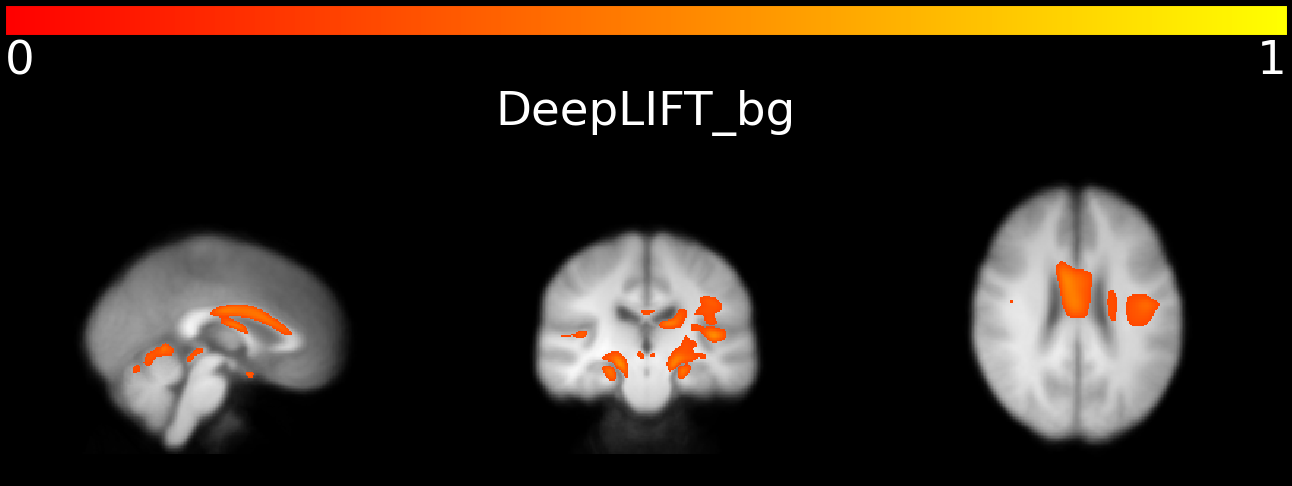}
    	            \caption{DeepLIFT$_\text{bg}$ T1R}
    	            \label{fig:DeepLIFT_overlay}
    	        \end{subfigure}
    	        \begin{subfigure}[b]{0.45\textwidth}
    	            \centering
    	            \includegraphics[width=\linewidth]{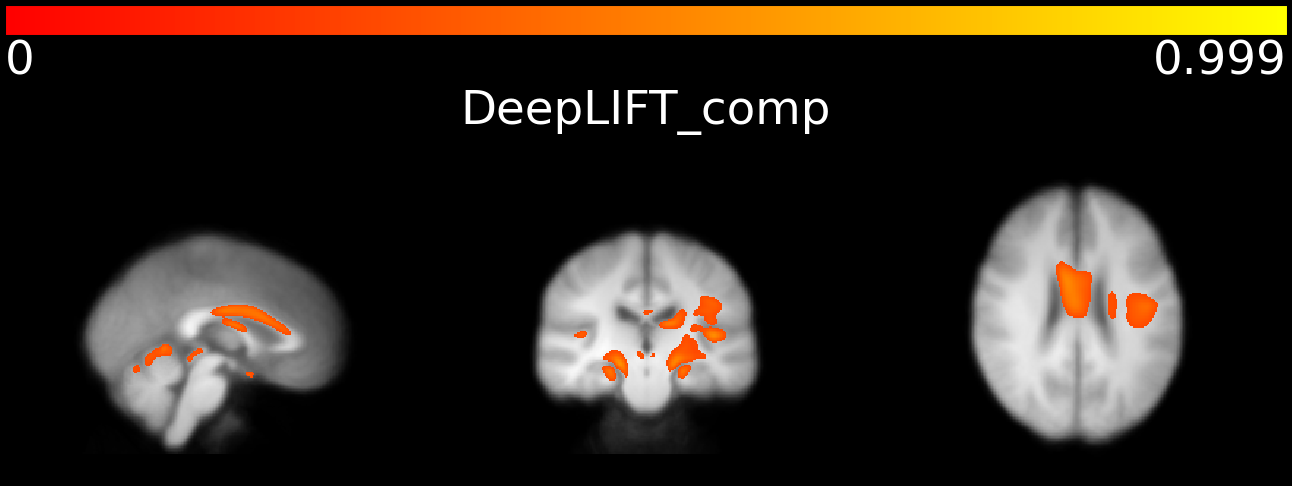}
    	            \caption{DeepLIFT$_\text{comp}$ T1R}
    	            \label{fig:DeepLIFT_comp_overlay}
    	        \end{subfigure}
                \begin{subfigure}[b]{0.45\textwidth}
    	            \centering
    	            \includegraphics[width=\linewidth]{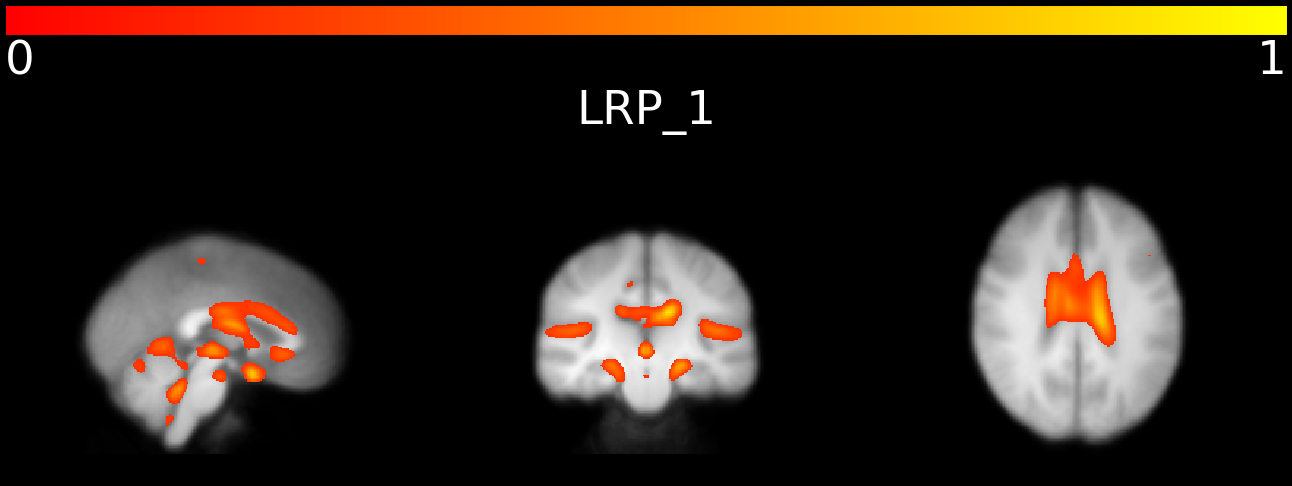}
    	            \caption{LRP$_1$ T1R}
    	            \label{fig:LRP_1_overlay}
                \end{subfigure}
                ~
                \begin{subfigure}[b]{0.45\textwidth}
                    \centering
                    \includegraphics[width=\linewidth]{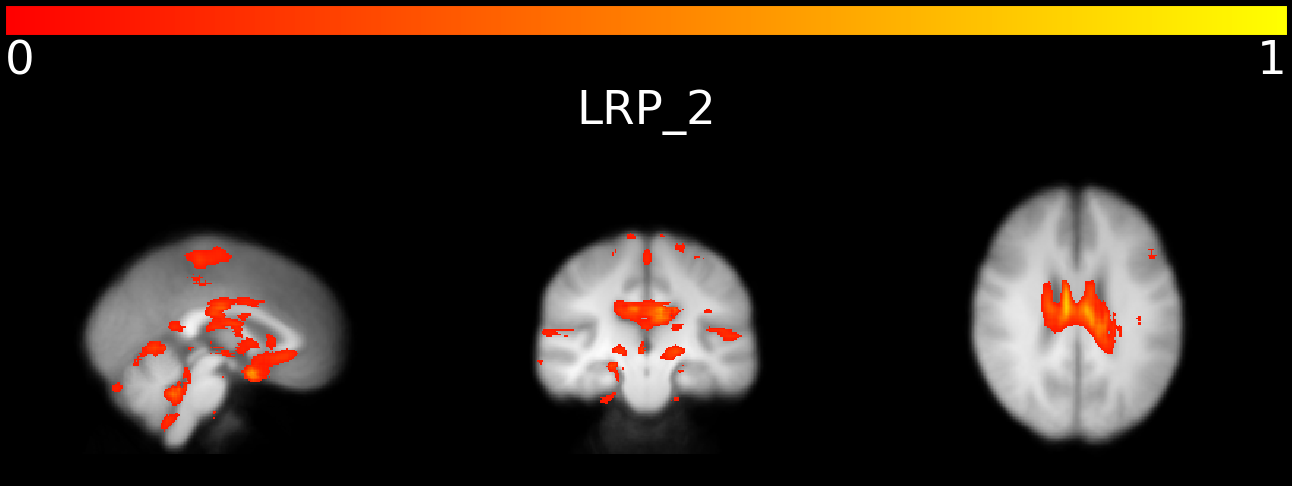}
                    \caption{LRP$_2$ T1R}
                    \label{fig:LRP_2_overlay}
                \end{subfigure}
                %~
                \begin{subfigure}[b]{0.45\textwidth}
                    \centering
                    \includegraphics[width=0.95\linewidth]{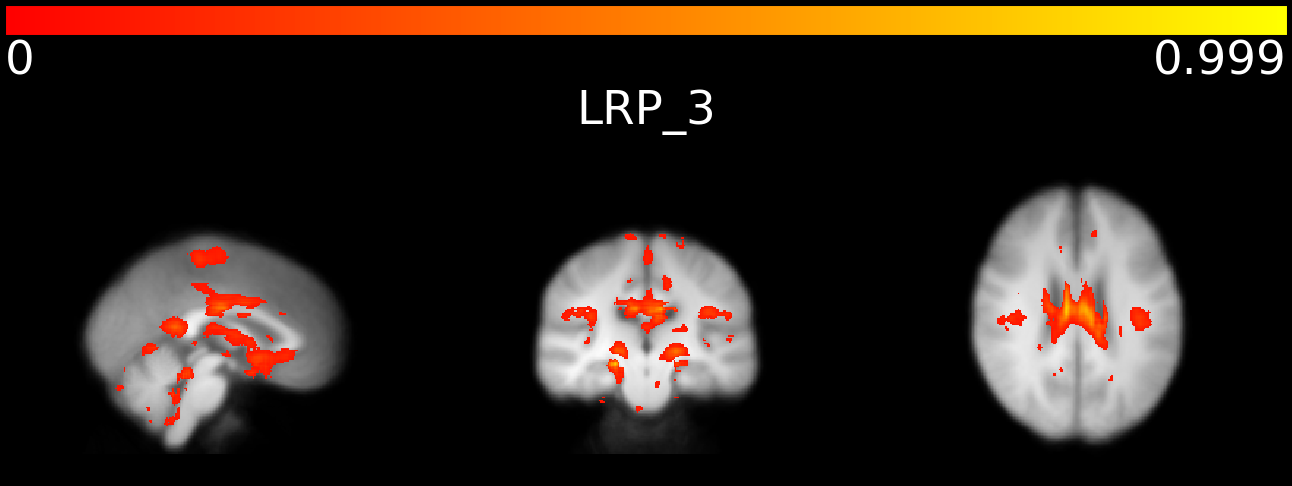}
                    \caption{LRP$_3$ T1R}
                    \label{fig:LRP_3_overlay}
                \end{subfigure}
                
                \caption{Sections of T1R for each method, overlaid on the composite MRI volume.}
                \label{fig:saliency_overlay}
            \end{figure}

        	We report here the findings of the saliency mapping tasks. Figure (\ref{fig:saliency_overlay}) shows sections of the composite volume overlaid with aggregate maps for each method. The composite volume was created as the aggregate of all the brain volumes from the test set, and each overlaid saliency map is similarly the aggregate of all the saliency maps from the test set for that method. The aggregate map was thresholded to the top-$1\%$ of voxel values to show the average distribution of T1R over the test set. We also show in Figure (\ref{fig:60y_m_overlay}) in Appendix (\ref{app:saliency_maps}) an example of overlaid LRP salience on sections of the volume of a 60-year-old male subject.
        
        	The permutation-based t-tests showed that there were statistically significant ($p<0.001$) differences between the saliency maps and the input volumes for each method in almost the entire brain volume.
	    
    	    \subsubsection{Regional Relevance Attribution Between Methods}
    	    
    	        Here we show the results of the assignment of relevance to brain regions by the five saliency mapping techniques. We compare the results to previous findings as well as to some of the medical BA literature, and examine differences between the methods.
    	        
    	        \begin{figure}[htbp]
                    \centering
                    \includegraphics[width=\linewidth]{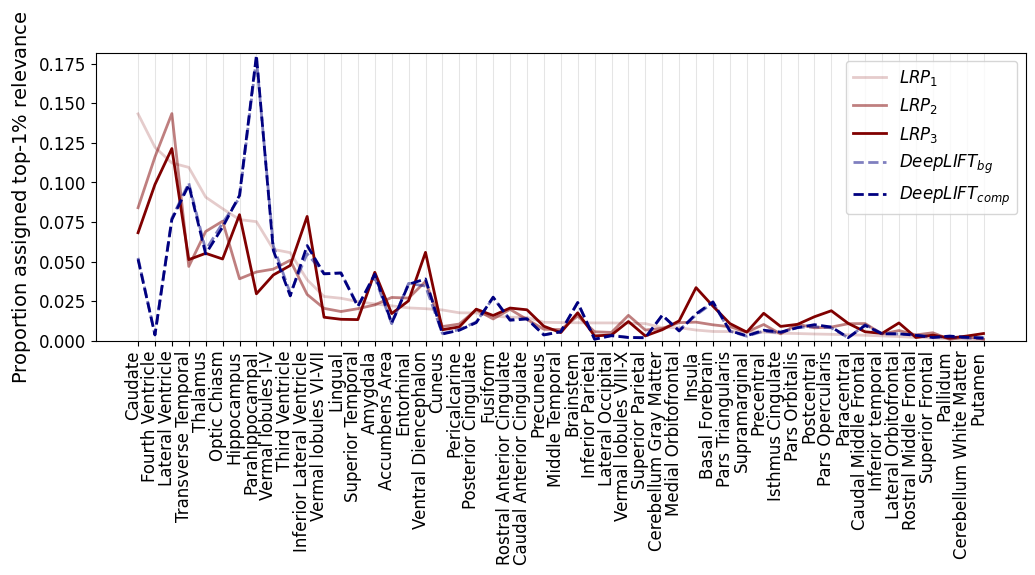}
                    \caption{Proportion of each region assigned Top-$1\%$ Relevance per method. Regions are ordered in descending proportion of T1R as determined by LRP$_1$.}
                    \label{fig:correspondence_ranking}
                \end{figure}
    	        
    	        Figure (\ref{fig:correspondence_ranking}) shows the proportion of each region that is assigned T1R, for each method. The relevance assignment in this case was averaged between the two hemispheres, since there was a large degree of symmetry in the distribution between hemispheres (see Appendix (\ref{app:saliency_maps})). The ordering of the regions along the x-axis is in descending order of proportion for LRP$_1$.
                
                It is clear that the LRP methods assign high T1R to the ventricles (particularly the lateral and fourth), which agrees not only with the medical literature \cite{Anderton2002, Raz2006}, but also with the findings of Levakov et al.\cite{Levakov2020} and Hofmann et al.\cite{Hofmann2021}. The DeepLIFT methods also assign high relevance to the lateral and inferior lateral ventricles, but less so than the LRP methods. Upon inspection of the DeepLIFT multipliers, it was clear that the same trend was present regardless of multiplication (masking) by the input. It appears that instead of assigning relevance to the ventricles themselves like the LRP methods, DeepLIFT tends to assign relevance to the immediate surrounds of the ventricles. DeepLIFT appears to highlight the areas which contract with age, whereas the LRP methods highlight the ventricles themselves, which dilate over time. This trend is displayed in the saliency sections of Figure (\ref{fig:saliency_overlay}).
                
                The transverse temporal gyrus was consistently assigned significant relevance by all methods, as were many limbic structures, such as the caudate nucleus,  hippocampus,  thalamus,  parahippocampal gyrus (most notably by DeepLIFT), and  diencephalon. Relevance of the limbic system to ageing has been shown by Gunbey et al. \cite{Gunbey2014}. Both methods tend to assign relevance to grey-matter-dense areas such as the vermal lobules of the cerebellum. This agrees with findings that grey matter density decreases with age, starting from late adolescence \cite{Peters2006, Sowell2003, Raz2006}. On the other hand, the cerebellum white matter tended not to be assigned much relevance. Although white matter lesions can be indicative of BA \cite{Peters2006, Anderton2002, Sowell2003, Raz2006}, these are not shown distinctively in T1-weighted images, and so we would not expect white matter regions to be assigned high relevance by this model. The optic chiasm tended to be assigned significant relevance. This may be due to its small size in conjunction with its proximity to other highly relevant structures such as the interpeduncular cistern \cite{Levakov2020}. The large amount of relevance assigned to the temporal gyrus agrees with findings by Lutz et al. \cite{Lutz2007} of involvement with brain ageing, and this relevance may be due to the degradation with age of the auditory cortex.
                
                There was a high degree of consistency in the relevance assignments within the LRP methods, and an even higher degree of consistency between the two DeepLIFT methods. Figure (\ref{fig:correspondence_ranking}) shows that the curves of the two DeepLIFT methods almost perfectly overlap. There was some similarity in the trend of relevance assignment between DeepLIFT and LRP. The biggest difference was that DeepLIFT assigns higher relevance to limbic structures like the parahippocampal gyrus, while LRP assigns more relevance to the ventricles.
        
            \subsubsection{Regional Relevance Attribution Based on BAG}
            
                Here we compare the distribution of relevance in individuals with small BAG to those with large BAG, both older and younger. We examine how regional relevance associates with BAG for older and younger individuals.
                
                \begin{figure}[H]
                    \centering
                    \includegraphics[width=\linewidth]{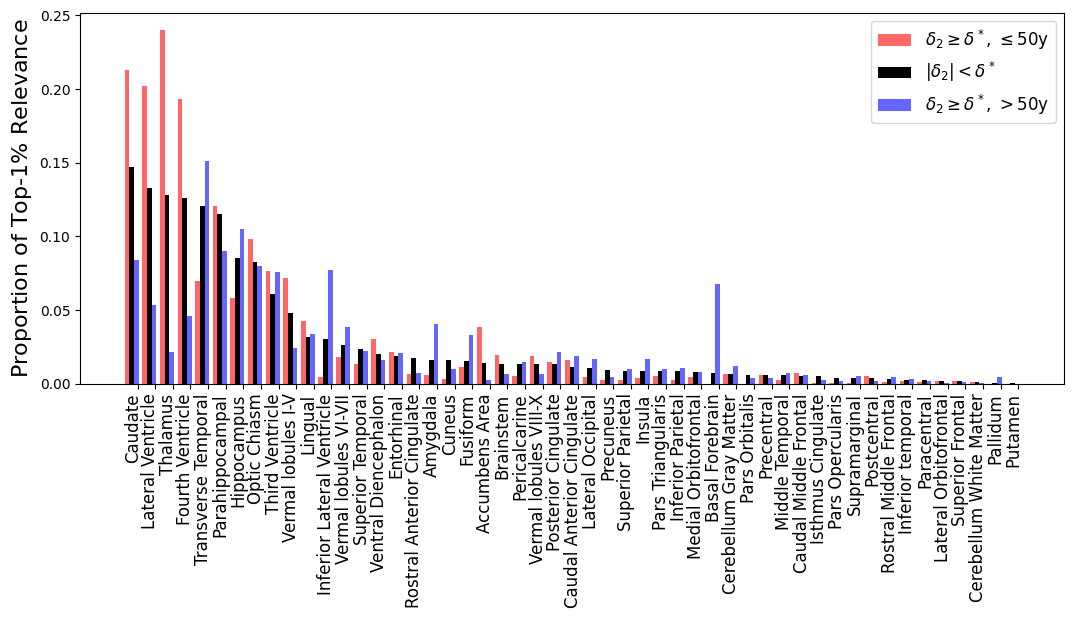}
                    \caption{Distributions of Top-$1\%$ Relevance, via LRP$_1$, in young ($\leq50$y) individuals with high BAG ($\delta_2>\delta^*$), elderly individuals ($>50$y) with high BAG, and individuals with small-to-moderate BAG ($|\delta_2|<\delta^*$). Regions are ordered by descending proportion of T1R in the small-to-moderate BAG group.}
                    \label{fig:LRP_1_dba_body}
                \end{figure}
                
                Figure (\ref{fig:test_dba_hist}) in Appendix (\ref{app:dba}) shows the distributions of the quantities $\delta_1$ and $\delta_2$, as defined in \cite{Smith2019}, for the test set. The correction from orthogonalisation to age shifted the distribution to become approximately Gaussian. This allowed us to threshold BAG reliably to compare BAG-salient regions in younger and older groups. On the test set, our threshold value was $\delta^*=\sigma=11.58$y.
                
                Figures (\ref{fig:DeepLIFT_bg_dba})-(\ref{fig:LRP_3_dba}) in Appendix (\ref{app:dba}) compare the distributions of T1R for these groups according to each method. We have included in Figure (\ref{fig:LRP_1_dba_body}) this distribution according to LRP$_1$. We see that several regions displayed significant changes in T1R assignment according to grouping by BAG and age. By way of example from Figure (\ref{fig:LRP_1_dba_body}), for large BAG values, DeepLIFT assigned to the fourth ventricle a significantly greater relevance in younger individuals than in older individuals. The opposite was true for the transverse temporal gyrus. Since increased relevance is predictive of higher BAG, these results suggest that markers of BAG change through the course of ageing. It appears that regions can be more indicative of BAG at some ages than others.
                
                \begin{figure}[htbp]
                    \centering
                    
                    \begin{subfigure}[b]{0.45\textwidth}
                        \centering
                        \includegraphics[width=\linewidth]{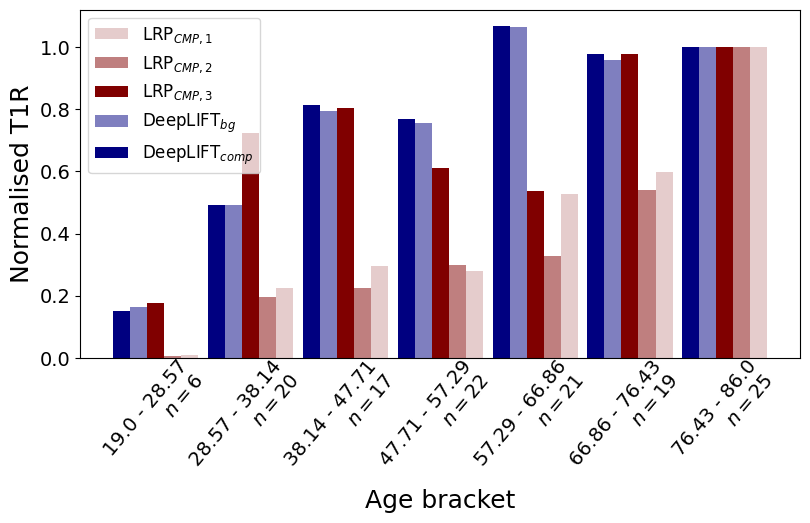}
                        \caption{Right Transverse Temporal Gyrus. Large Standard Deviation in T1R over age brackets, and relevance increases with age.}
                        \label{fig:brackets_transverse_temporal}
                    \end{subfigure}
                    \hspace{3mm}%
                    \begin{subfigure}[b]{0.45\textwidth}
                        \centering
                        \includegraphics[width=\linewidth]{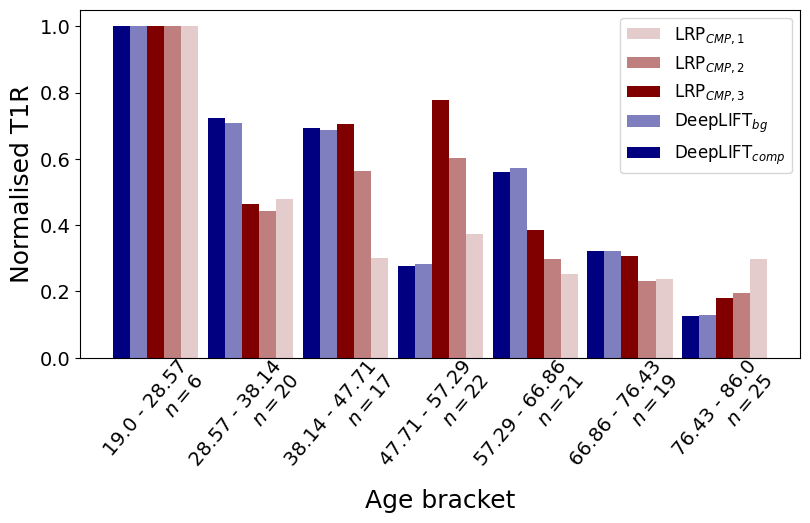}
                        \caption{Right Fourth Ventricle. Large Standard Deviation in T1R over age brackets, and relevance decreases with age.}
                        \label{fig:brackets_fourth_ventricle}
                    \end{subfigure}
                    \par\bigskip %
                    \begin{subfigure}[b]{0.45\textwidth}
                        \centering
                        \includegraphics[width=\linewidth]{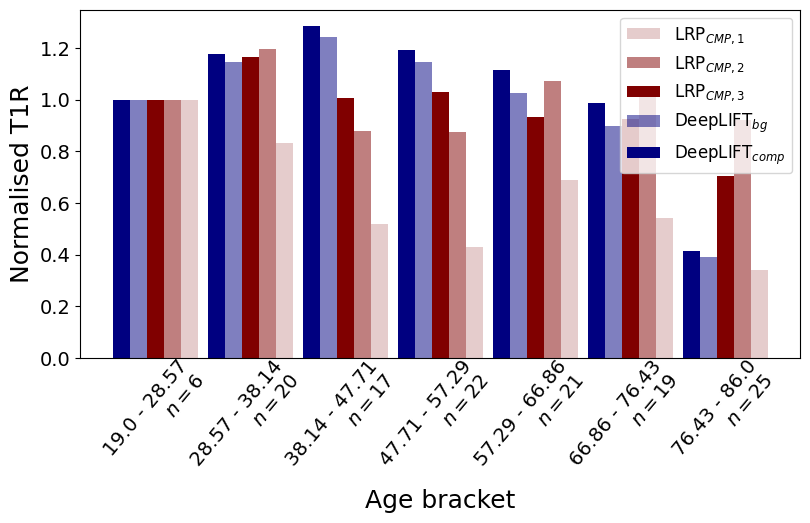}
                        \caption{Right Lateral Ventricle. Low CoV in T1R over age brackets.}
                        \label{fig:brackets_lateral_ventricle}
                    \end{subfigure}
                    \hspace{3mm}%
                    \begin{subfigure}[b]{0.45\textwidth}
                        \centering
                        \includegraphics[width=\linewidth]{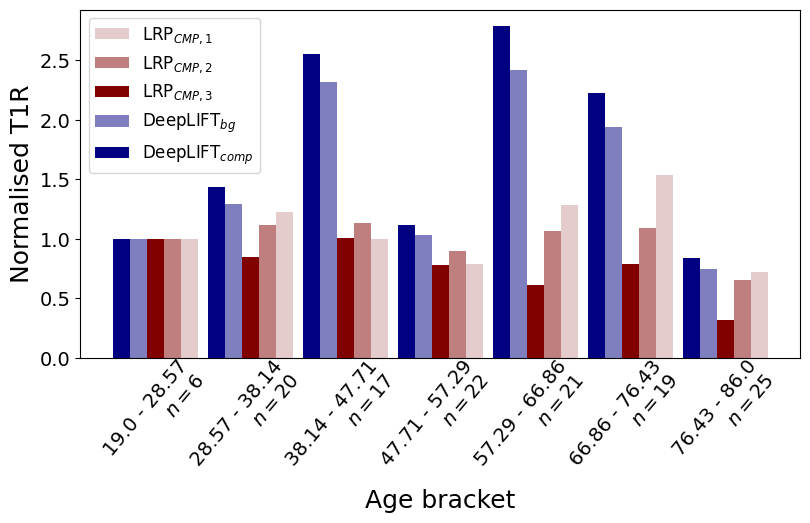}
                        \caption{Right Caudate Nucleus. Low CoV in T1R over age brackets.}
                        \label{fig:brackets_caudate}
                    \end{subfigure}
                    
                    \caption{`Proportion' of Top-$1\%$ Relevance per method over the age brackets. The proportion of T1R is normalised in each sub-figure such that either the youngest or oldest group have T1R assignment of 1. Brackets all have an age range of $9.57$y. Four example regions are shown.}
                    \label{fig:age_brackets}
                \end{figure}
	    
	        \subsubsection{Regional Relevance Attribution Across Age Brackets}
            
                Here we examine the change in relevance across age brackets for brain regions according to each saliency mapping method. In Figure (\ref{fig:sd_and_cov}) in Appendix (\ref{app:age_brackets}) we show the standard deviations and coefficients of variation in the proportion of regions assigned T1R over all age brackets.
                
                One region which has a large SD over ages for all methods is the transverse temporal gyrus. We show the distribution across age brackets for the right hemispheric structure in Figure (\ref{fig:brackets_transverse_temporal}) for each method. For uniformity in scale we normalise the proportions per method such that either the oldest or youngest age bracket is $1$ for all methods. All methods, especially LRP, show an increase in relevance of this region with greater age.
                
                Greater attribution of relevance in older subjects than in younger subjects suggests that the structure bears more information about BA and BAG in older individuals. This agrees with our findings about relevance distribution as a function of BAG. The transverse temporal gyrus for example is a more salient indicator of accelerated brain ageing in older subjects than in younger subjects. On the other hand, we see in Figure (\ref{fig:brackets_fourth_ventricle}) the distribution across age brackets of the same quantity for the right fourth ventricle. This region too is shown to have significant SD across age brackets. There is a marked decline in assignment of T1R from younger to older ages. This suggests that while the fourth ventricle is universally relevant to BA and BAG, it is most telling of ageing in younger individuals.
                
                Figure (\ref{fig:brackets_lateral_ventricle}) shows the distribution across age brackets of the same quantity, now for the right lateral ventricle. This was shown to have uniformly low CoV across the methods. We see that there is a decrease in relevance in the oldest age bracket for DeepLIFT and LRP$_3$, but otherwise the relevance assignment is uniform.
                
                Figure (\ref{fig:brackets_caudate}) shows the age bracket distribution of T1R for the right caudate nucleus. The region has a low CoV for each method. Although there is some disparity in assigned relevance over the age brackets (especially within DeepLIFT relative to the others), there is no overarching upward or downward trend.
                
	\section{Discussion and Conclusions}
	
	    While the test set MAE achieved by the model of $6.55$y is not SOTA, no BA model in the literature has achieved such accuracy on a dataset of comparable size. This would indicate that that while very large datasets may be necessary to achieve SOTA performance, they may not be necessary simply to train a BA regression model of reasonable accuracy (with a coefficient of correlation to chronological age of $r=0.89$ -- see Figure (\ref{fig:test_regression}) of Appendix (\ref{app:regression_model})). Since it can be difficult to access very large datasets of healthy individuals across the full adult lifespan, it may be preferable for some studies to use smaller datasets such as Cam-CAN. Table (\ref{tab:study_maes}) of Appendix (\ref{app:regression_model}) compares the MAE performance of several high-performing BA regression models and shows the dataset size used in each study.
	    
	    The three LRP methods are relatively consistent in their distribution of T1R among the brain regions, and the two DeepLIFT functions even more so. By way of comparison, there is a notable similarity between LRP and DeepLIFT in BA relevance assignment. This is to be expected, since the methods are very closely related \cite{Shrikumar2016}, and DeepLIFT can in fact represent a generalisation of some LRP methods \cite{Ancona2018}. The LRP methods assigned T1R to the ventricles in large proportion, agreeing with the findings of Levakov et al. \cite{Levakov2020} and Hofmann et al. \cite{Hofmann2021}. DeepLIFT, however, tends to assign relevance to the immediately surrounding regions, and more heavily to limbic areas like the parahippocampal gyrus. There is no precedent for such differences in explanations in the literature, past suggestions for what such differences may mean, or where they may come from. We speculate that LRP and DeepLIFT may highlight the importance of the same phenomenon from two different perspectives, in that the dilation of the ventricles and the atrophy of the surrounding regions are causally coupled.
        
	    Both LRP and DeepLIFT assign high relevance to the ventricles and to limbic structures, which agrees both with our expectations and with literature on BA \cite{Anderton2002, Raz2006, Lutz2007, Gunbey2014}. High proportions of relevance are assigned to grey-matter-dense areas, which also agrees with some medical literature \cite{Anderton2002, Raz2006, Gunbey2014}. 
	    
	    There are non-trivial differences in the distributions of relevance between LRP and DeepLIFT, but both methods highlight known areas of importance to brain ageing. The utility of one method over the other may depend on the user's main regions of interest. It may be best to use a combination of LRP and DeepLIFT to form BA explanations. LRP$_1$ is the most commonly used LRP method, and is the least noisy of those we implemented, since higher values of $\alpha$ create greater contrast in the saliency maps \cite{Bach2015, Grigorescu2019}. This is illustrated in Figures (\ref{fig:LRP_1_overlay})-(\ref{fig:LRP_3_overlay}), where higher values of $\alpha$ decrease the smoothness of T1R clusters. As we have seen, the two DeepLIFT methods perform extremely similarly, and both are simple to implement; thus we cannot conclude that one is more useful for BA explanation than the other.
	    
	    We found that large BAG values are associated with distributions in relevance among brain regions that can vary to a large degree depending on age. This suggests that the degree of saliency of a region towards BAG is age-dependent. Upon inspection of age-bracket distribution of relevance to brain regions, we showed that a tripartite pattern emerges. Some regions increase in relevance with age, some decrease with age and others retain relatively uniform relevance with age. From our findings with BAG, decreases with age would suggest that the region is more informative of BAG in younger individuals than in older individuals (for example, the fourth ventricle). Similarly increases with age would suggest that the region is more informative of BAG in older individuals than in younger ones (for example, the transverse temporal gyrus). Uniformity across ages in relevance would suggest that the region is consistently informative of BAG (for example, the caudate nucleus).
	    
	    This tripartite pattern of trajectories was unexpected, since previous studies have only reported on increased relevance with age \cite{Hofmann2021}. Indeed, many regions decrease in the proportion assigned T1R with age. This can be explained for a structure like the right fourth ventricle, by the fact that although the structure is informative of BA at all ages, it is expected to dilate with age. A younger individual with dilated ventricles is clearly experiencing accelerated brain ageing.
	    
	    These findings may be used as baseline regional trajectories for BA salience in a clinical setting. As part of a DL pipeline for BA regression, clinicians can create saliency maps for patients' BA predictions and compare the regional distributions of T1R to these baselines. Large individual BAG can be assessed regionally by these methods. Saliency mapping can be performed post-hoc on the regression model in close to real-time, so as to have BA predictions accompanied by an explanation in a clinical setting.
	    
	    Without access to this model and its accompanying results, clinicians will still be able to use the underlying methods -- perhaps with more powerful models trained on larger datasets -- to create and compare relevance trajectories for clinical use.
	    
	    In addition to the BA regression model and explanation pipeline, and the baseline regional saliency trajectories, we have also contributed a regional analysis of BA saliency in individuals with accelerated BA (large BAG), showing that some regions show increased BA in younger individuals and others show increased BA in older individuals. Finally, we have also contributed a comparison of several saliency mapping techniques from two classes of technique, showing that there are some significant differences in explanations between the classes of explanation techniques.
	    
	    A limitation of this work is the use of single-fold cross-validation in the analysis of the BA regression model. A 5-fold cross-validation applied to larger datasets would ensure robust model accuracy on unseen data. In the future it would also be of great interest to examine how the distribution of relevance within the brain may change according to different model architectures and across different datasets. Another potential avenue for future work is to see how the size of a dataset may influence the distribution of relevance through the brain volume. It would also be of interest to compare the saliency maps of other, possibly more computationally expensive techniques, such as Integrated Gradients \cite{Sundararajan2016, Sundararajan2017}.
        
        % Acknowledgments---Will not appear in anonymised version
        
        %%%%%%%%%%%%%%%%%%
        %ACKNOWLEDGEMENTS%
        %%%%%%%%%%%%%%%%%%
        \section*{Acknowledgements}
            
            \scriptsize
            Funding for this work was provided by the University of Cape Town Postgraduate Funding Office, the University of Cape Town Computer Science Department, and the ETDP-SETA. Training and saliency mapping was performed on the Lambda Cloud GPU platform, using their A6000 instances. Preprocessing utilised FSL's \cite{Smith2004} BET and FLIRT tools \cite{Smith2002, Jenkinson2001, Jenkinson2002}. The randomise tool was used as well \cite{Winkler2014}, for the permutation-based t-tests. Brain regions were determined using the CerebrA 2009c atlas \cite{Collins1999, Fonov2009, Manera2020}.
            \normalsize
    
    \newpage
    
    \appendix
    
    \renewcommand\thefigure{\thesection.\arabic{figure}}
    \renewcommand\thetable{\thesection.\arabic{table}}
    
    \section{Pre-processing}\label{app:preprocessing}
        
        \setcounter{figure}{0}
        \setcounter{table}{0}
        
        The standard pre-processing of the Cam-CAN dataset is detailed by Taylor et al. \cite{Taylor2017}. We used FSL's Brain Extraction Tool (BET) \cite{Smith2002} to perform skull-stripping on the T1-weighted volumes, using the recommended fractional intensity of $0.5$. We then aligned all volumes to MNI space by registration to a standard MNI volume \cite{Manera2020}, using the FLIRT tool \cite{Jenkinson2001, Jenkinson2002}. This was done primarily for spatial correspondence with the region atlas \cite{Collins1999, Fonov2009}. Voxel values were normalised using Global Contrast Normalisation (GCN) to have unit variance within each volume.
    
    \section{Regression Model}\label{app:regression_model}
        
        \setcounter{figure}{0}
        \setcounter{table}{0}
        
\begin{figure}[htbp]
            \begin{floatrow}
                \ffigbox{\includegraphics[width=\linewidth]{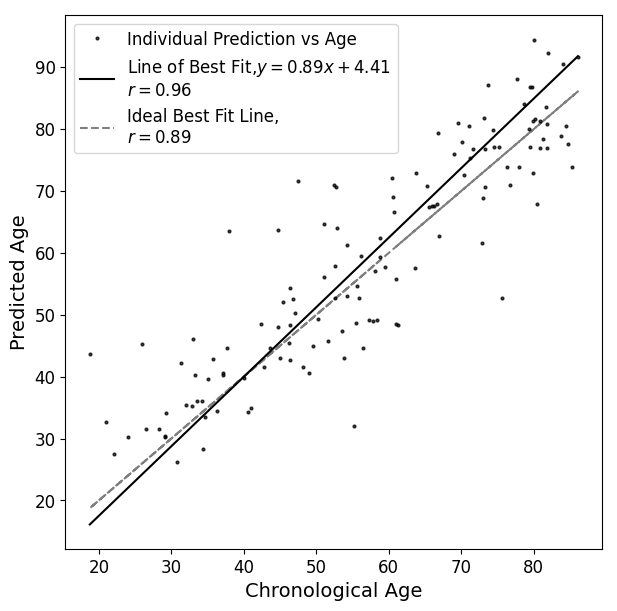}}{\caption{Test set regression plot}\label{fig:test_regression}}
                \capbtabbox{%
                    \begin{tabular}{ll}
                        \hline
                        \bfseries Metric & \qquad\bfseries Result\\
                        \hline
                        MAE & \qquad6.5457\\
                        MSE & \qquad72.5481\\
                        MAPE & \qquad13.5305
                    \end{tabular}
                }{\caption{Metrics}\label{tab:model_metrics}}
            \end{floatrow}
        \end{figure}

        Figure (\ref{fig:test_regression}) shows the regression plot for our ResNet model on the test set, and Table (\ref{tab:model_metrics}) shows the performance metrics. We note that the MAE of $6.5$y is far from state-of-the-art, but also that the size of the dataset is far less than those of previous works. Table (\ref{tab:study_maes}) shows the performances of some key studies for BA regression, with the number of individuals in the datasets used.
        
        Our ResNet had filter sizes $\left[32, 32, 64, 64, 128\right]$ in the main branches of the five residual modules. This configuration was adapted from the lightweight model of Jonsson et al. \cite{Jonsson2019}.
        
        \begin{table}[htbp]
            \begin{tabular}{|l|l|l|l|l|l|l|l|}
                \hline
                \textbf{Author}                                                               & Ours & \begin{tabular}[c]{@{}l@{}}Cole\\ et al.\\(2017) \end{tabular} & \begin{tabular}[c]{@{}l@{}}J\'onsson\\ et al.\\(2019) \end{tabular} & \begin{tabular}[c]{@{}l@{}}Kossaifi\\  et al.\\(2020) \end{tabular} & \begin{tabular}[c]{@{}l@{}}Levakov\\ et al.\\(2020) \end{tabular} & \begin{tabular}[c]{@{}l@{}}Hofmann\\ et al.\\(2021) \end{tabular} & \begin{tabular}[c]{@{}l@{}}Peng\\ et al.\\(2021) \end{tabular} \\ \hline
                \textbf{Dataset Size}                                                         & 656  & 2001                                                           & 12378                                                                              & 19100                                                              & 10176                                                             & 2016                                                              & 14503                                                          \\ \hline
                \textbf{\begin{tabular}[c]{@{}l@{}}Test MAE (y)\\ on T1 Volumes\end{tabular}} & 6.55\qquad\qquad & 4.65\qquad                                                           & 4.00\qquad                                                                                & 2.69\qquad                                                               & 3.02\qquad                                                              & 3.95\qquad                                                              & 2.14\qquad                                                           \\ \hline
            \end{tabular}
            \caption{Test MAE performances of several key BA regression studies, as compared to ours. Our dataset is at least three times smaller than any other used here.}
            \label{tab:study_maes}
        \end{table}
    
    \section{Saliency Maps}\label{app:saliency_maps}
        
        \setcounter{figure}{0}
        \setcounter{table}{0}
        
        LRP and DeepLIFT are relevance decomposition methods which assign scores to the elements of a model's input space according to their saliency to the model's decision. Both methods take into account the learned parameters of the model and the activations at each layer of the model specific to the inference of a given input. Relevance decomposition refers to the backward propagation of relevance from the output layer of a model to its input. We chose LRP and DeepLIFT because these are both fast and computationally inexpensive saliency mapping methods. Both methods have also been used before in reasoning for brain imaging \cite{Eitel2019, Grigorescu2019, Hofmann2021, Pianpanit2019, Gupta2019}, and offer directionality in their explanations, unlike simpler method such as SmoothGRAD \cite{Smilkov2017}).
        
        The decomposition rules for LRP$_\text{CMP}$ \cite{Kohlbrenner2019} from a layer $(L + 1)$ to a layer $L$ are as follows:
        
        For fully-connected layers, the $\varepsilon$-rule \cite{Bach2015} is used:
        \begin{equation}
            R_i^{\left(L\right)} = \sum_j{\frac{z_{ij}}{\sum_{i'}{z_{i'j} + \varepsilon}}R^{\left(L + 1\right)}_j}
        \end{equation}
        where $R_i^{\left(L\right)}$ is the relevance assigned to neuron $i$ in layer $L$, $z_{ij}=x_iw_{ij}$ is the product of previous layer activations and the learned weights between the layers, and $\varepsilon$ is a small constant for numerical stability.
        
        For convolutional, BatchNorm and pooling layers (apart from the input layer), the LRP$_{\alpha\beta}$ rule \cite{Bach2015} is used:
        \begin{equation}
            R_i^{(L)} = \sum_j\left(\alpha\dfrac{(x_iw_{ij})^+}{\sum_{i'}(x_iw_{ij})^++b_j^+} - \beta\dfrac{(x_iw_{ij})^-}{\sum_{i'}(x_iw_{ij})^-+b_j^-}\right)R_j^{(L + 1)}
        \end{equation}
        with the constraint that $\alpha - \beta = 1$. $w_{ij}^+$ refers to the positive parts only of the weights and $b_j^+$ to the positive parts only of the biases.
        
        For the input layer, the $z^\mathcal{B}$-rule \cite{Montavon2017} is used:
        \begin{equation}
	        R_i^{(L)} = \sum_j\frac{z_{ij} - l_iw_{ij}^+ - h_iw_{ij}^-}{\sum_{i'}z_{i'j} - l_{i'}w_{i'j}^+ - h_{i'}w_{i'j}^-}R_j^{(L + 1)}
	    \end{equation}
	    where $h_i$ and $l_i$ are the highest and lowest input values respectively.
	    
	    The parameters $\alpha$ and $\beta$ of LRP act as contrast parameters, with higher values of $\alpha$ giving higher contrast in the saliency maps \cite{Bach2015}.\par\bigskip
	    
	    The DeepLIFT decomposition rules from a neuron $y$ in a layer to a neuron $x$ in a layer above are given by contribution scores \cite{Shrikumar2017}, allocated according to layer-specific rules. The \textit{Linear Rule} for all linear layers, allocates contribution scores as follows:
	    \begin{align*}
	        &C_{\Delta x_i^+\Delta y^+} = 1\left\{w_i \Delta x_i > 0\right\}w_i \Delta x_i^+,\\
            &C_{\Delta x_i^-\Delta y^+} = 1\left\{w_i \Delta x_i > 0\right\}w_i \Delta x_i^-,\\
            &C_{\Delta x_i^+\Delta y^-} = 1\left\{w_i \Delta x_i < 0\right\}w_i \Delta x_i^+,\\
            &C_{\Delta x_i^-\Delta y^-} = 1\left\{w_i \Delta x_i < 0\right\}w_i \Delta x_i^-
	    \end{align*}
	    where $C_{x_iy}$ is the contribution of $x$ to $y$, $\Delta x$ refers to the difference-from-reference of the neuron $x$, and $\Delta x^+$ is the positive-only part of that quantity. The difference-from-reference for a neuron $x$ is calculated as the difference between the activations of that neuron for the given input and a reference input, $\Delta x = x^0 - x$.
	    
	    The \textit{Reveal-Cancel Rule} is used for non-linear layers and allocates contribution scores as follows:
	    \begin{align*}
	        C_{\Delta x^+\Delta y^+} =& \frac{1}{2}\left(f(x^0 + \Delta x^+) - f(x^0)\right)\\
            &+ \frac{1}{2}\left(f(x^0 + \Delta x^- + \Delta x^+) - f(x^0 + \Delta x^-)\right)\\
            C_{\Delta x^-\Delta y^-} =& \frac{1}{2}\left(f(x^0 + \Delta x^-) - f(x^0)\right)\\
            &+ \frac{1}{2}\left(f(x^0 + \Delta x^+ + \Delta x^-) - f(x^0 + \Delta x^+)\right)
	    \end{align*}
	    
	    where $f$ refers to the nonlinearity function -- in our case the Softplus function.
	    
	    The reference input is a choice of the user for DeepLIFT. The authors \cite{Shrikumar2017} state that there can be multiple `good' choices for reference inputs, and that the choice is dependent on the task. This is our reason for having chosen two different reference inputs to compare. Indeed, the results were almost identical for distribution of T1R. Ancona et al. \cite{Ancona2018} show that LRP and DeepLIFT are similar enough that DeepLIFT relevance decomposition can be rewritten to look like LRP, with difference-from-reference values. The authors note that the significant difference between the methods is the use of reference inputs in DeepLIFT.\par\bigskip
	    
\begin{figure}[htbp]
	        \centering
	        \includegraphics[width=0.7\linewidth]{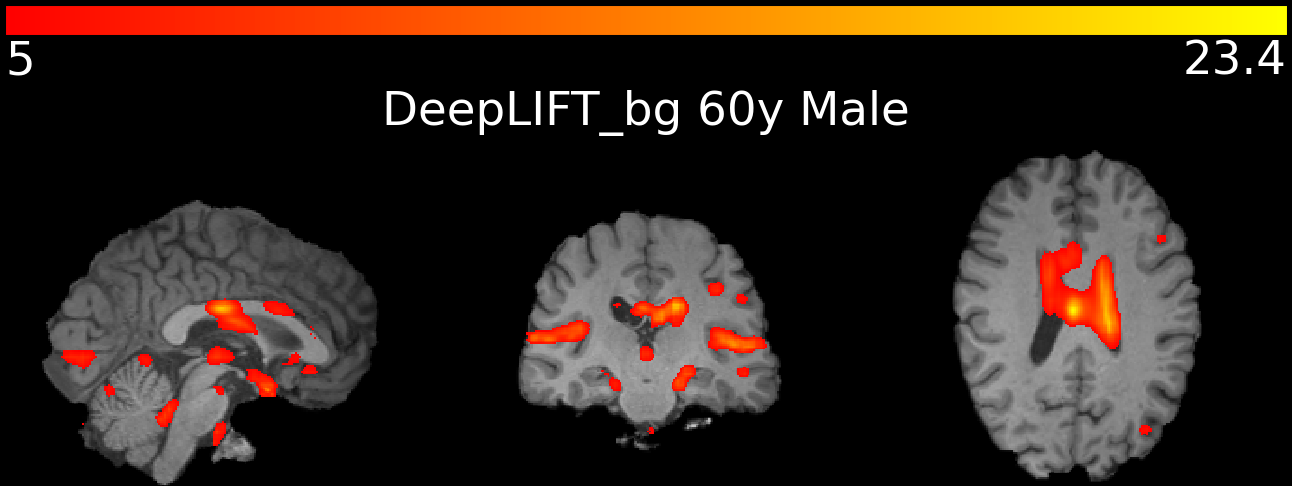}
	        \caption{Individual saliency map for a 60-year-old male subject, thresholded to T1R and overlaid onto the subject's MRI volume}
	        \label{fig:60y_m_overlay}
	    \end{figure}

	    We show in Figure (\ref{fig:60y_m_overlay}) the overlay of T1R of a single subject onto their specific brain volume, as an example.
	    
\begin{figure}[htbp]
            \centering
            
            \begin{subfigure}[b]{0.3\textwidth}
                \centering
                \includegraphics[width=\linewidth]{region_correlations/LRP_1_top_1_corr_his.png}
                \caption{LRP$_1$}
                \label{fig:LRP_1_corr_hist}
            \end{subfigure}%
            \begin{subfigure}[b]{0.3\textwidth}
                \centering
                \includegraphics[width=\linewidth]{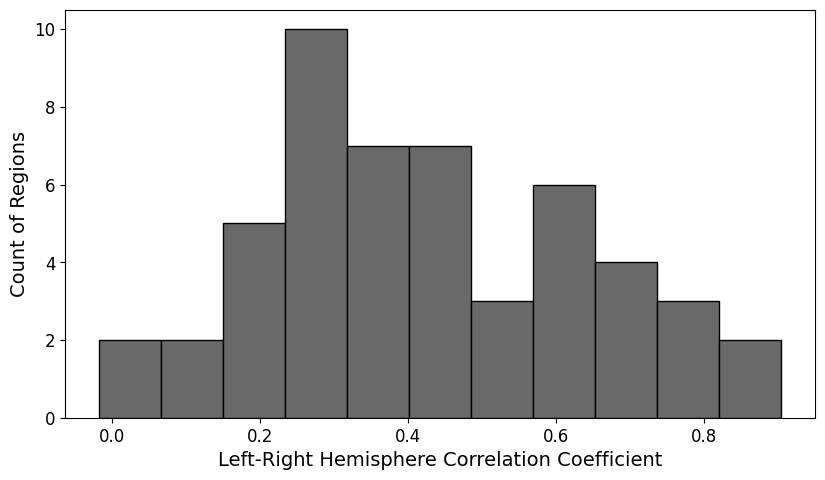}
                \caption{LRP$_2$}
                \label{fig:LRP_2_corr_hist}
            \end{subfigure}%
            \begin{subfigure}[b]{0.3\textwidth}
                \centering
                \includegraphics[width=\linewidth]{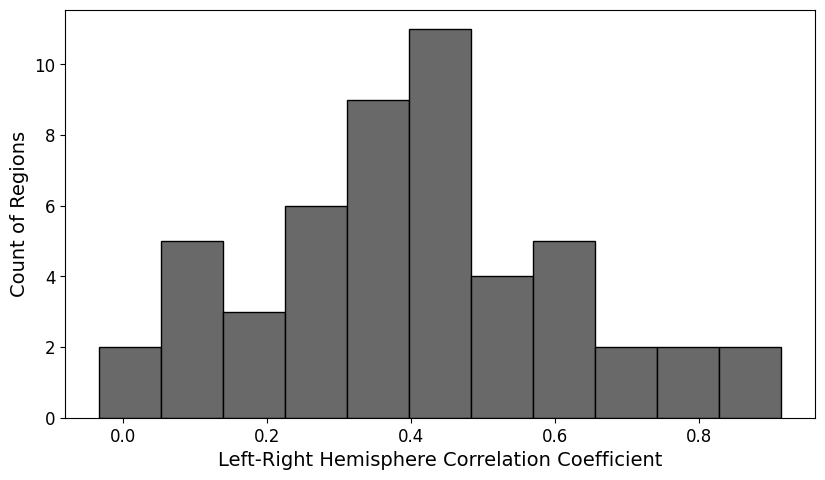}
                \caption{LRP$_3$}
                \label{fig:LRP_3_corr_hist}
            \end{subfigure}
            \begin{subfigure}[b]{0.3\textwidth}
                \centering
                \includegraphics[width=\linewidth]{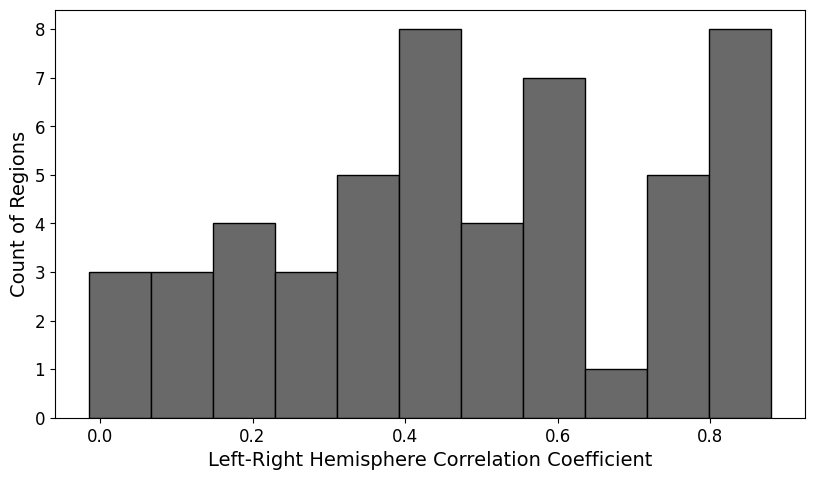}
                \caption{DeepLIFT$_\text{bg}$}
                \label{fig:DeepLIFT_bg_corr_hist}
            \end{subfigure}%
            \begin{subfigure}[b]{0.3\textwidth}
                \centering
                \includegraphics[width=\linewidth]{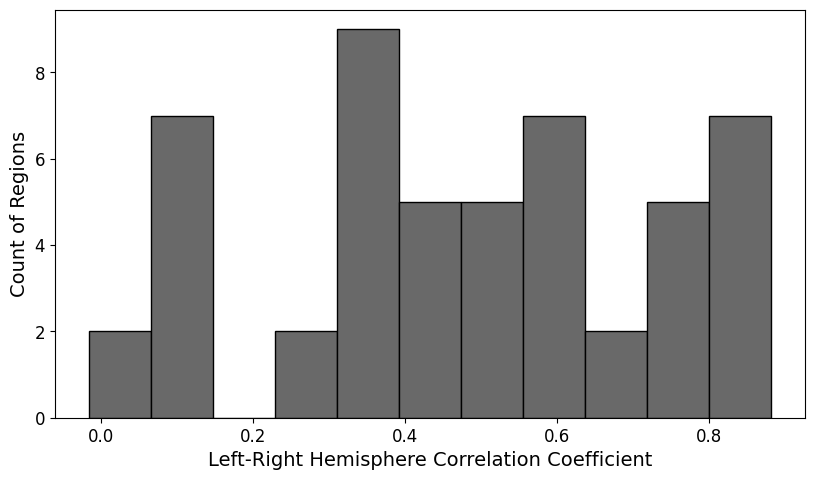}
                \caption{DeepLIFT$_\text{comp}$}
                \label{fig:DeepLIFT_comp_corr_hist}
            \end{subfigure}
            
            \caption{Histograms of correlation coefficients of T1R between left- and right-hemispheric structures.}
            \label{fig:correlation_hist}
        \end{figure}

        Figure (\ref{fig:correlation_hist}) shows the distributions of the coefficients of correlation for relevance attribution to the same region in opposite hemispheres, for each method. We see that the majority of regions have high coefficients of correlation between hemispheres. A few do not have high correlations, and these tend to be regions which are assigned low proportions of T1R. This correlation between hemispheres implies a symmetry across the hemispheres, and therefore allows us to average relevance between hemispheres in our analysis.
        
    \section{BAG}\label{app:dba}
        
        \setcounter{figure}{0}
        \setcounter{table}{0}
        
        \begin{figure}[H]
             \centering
             \includegraphics[width=\linewidth]{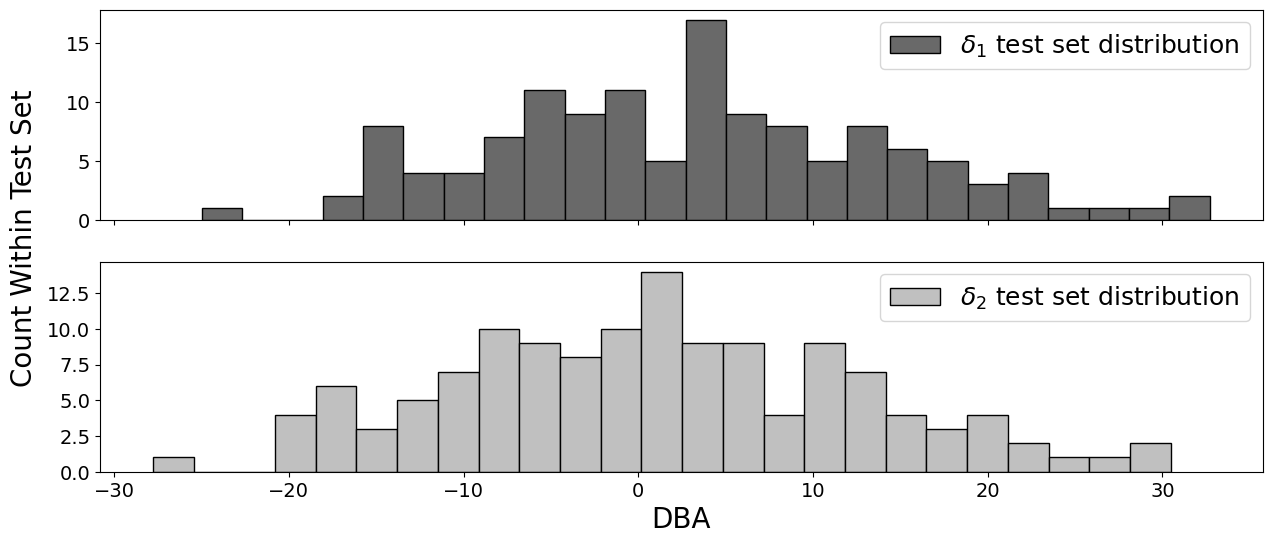}
             \caption{Distributions of BAG (sometimes called DBA) by $\delta_1$ and $\delta_2$ (as defined by \cite{Smith2019}) on the test set. The age-orthogonality correction of $\delta_2$ shifts this distribution to be centered approximately at 0 (mean moves from $3.30$y to $0.98$y), and distributed more evenly to either side, becoming approximately Normal (range shifts from $\left(-24.98, 32.70\right)$ to $\left(-27.79, 30.47\right)$).}
             \label{fig:test_dba_hist}
        \end{figure}
        
        \begin{figure}[H]
            \centering
            \includegraphics[width=0.95\linewidth]{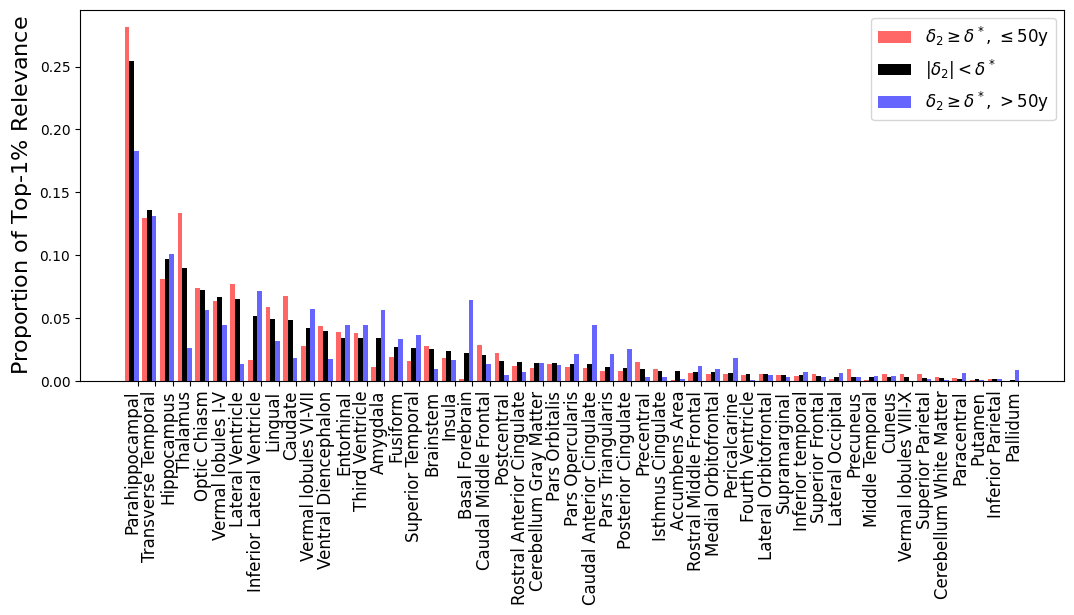}
            \caption{Distributions of Top-$1\%$ Relevance, via DeepLIFT$_\text{bg}$, in young ($\leq50$y) individuals with high BAG ($\delta_2>\delta^*$), elderly individuals ($>50$y) with high BAG, and individuals with small-to-moderate BAG ($|\delta_2|<\delta^*$). Regions are ordered by descending proportion of T1R in the small-to-moderate BAG group.}
            \label{fig:DeepLIFT_bg_dba}
        \end{figure}
        
        \begin{figure}[htbp]
            \centering
            \includegraphics[width=0.95\linewidth]{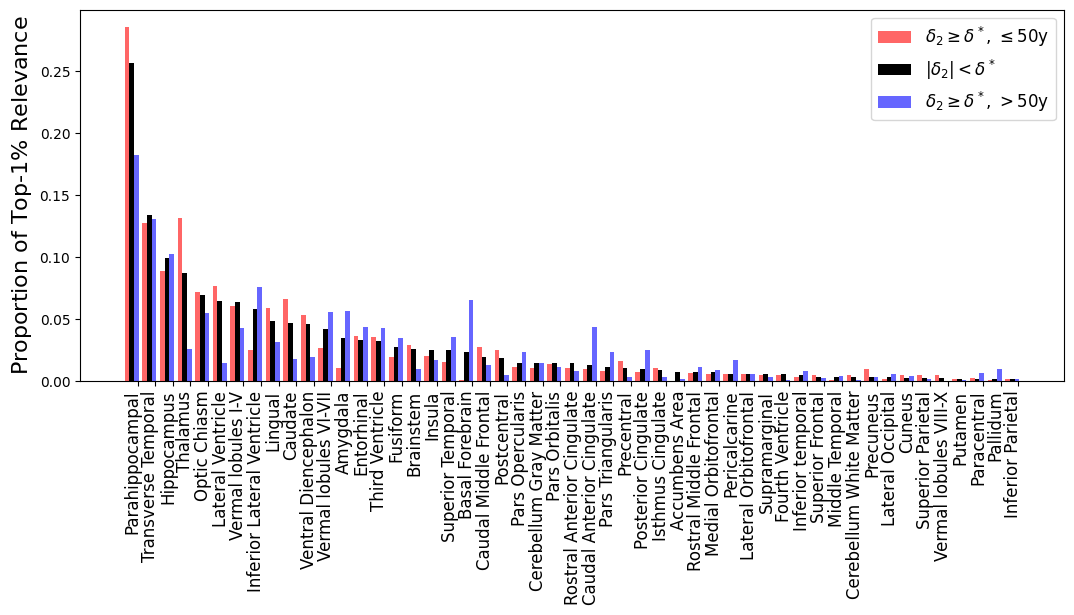}
            \caption{Distributions of Top-$1\%$ Relevance, via DeepLIFT$_\text{comp}$, in young ($\leq50$y) individuals with high BAG ($\delta_2>\delta^*$), elderly individuals ($>50$y) with high BAG, and individuals with small-to-moderate BAG ($|\delta_2|<\delta^*$). Regions are ordered by descending proportion of T1R in the small-to-moderate BAG group.}
            \label{fig:DeepLIFT_comp_dba}
        \end{figure}
        
        %\begin{figure}[H]
        %    \centering
        %    \includegraphics[width=0.95\linewidth]{outlier_graphs/LRP_1_regions_dba.png}
        %    \caption{Distributions of Top-$1\%$ Relevance, via LRP$_1$, in young ($\leq50$y) individuals with high BAG ($\delta_2>\delta^*$), elderly individuals ($>50$y) with high BAG, and individuals with small-to-moderate BAG ($|\delta_2|<\delta^*$). Regions are ordered by descending proportion of T1R in the small-to-moderate BAG group.}
        %    \label{fig:LRP_1_dba}
        %\end{figure}
        
        \begin{figure}[H]
            \centering
            \includegraphics[width=0.95\linewidth]{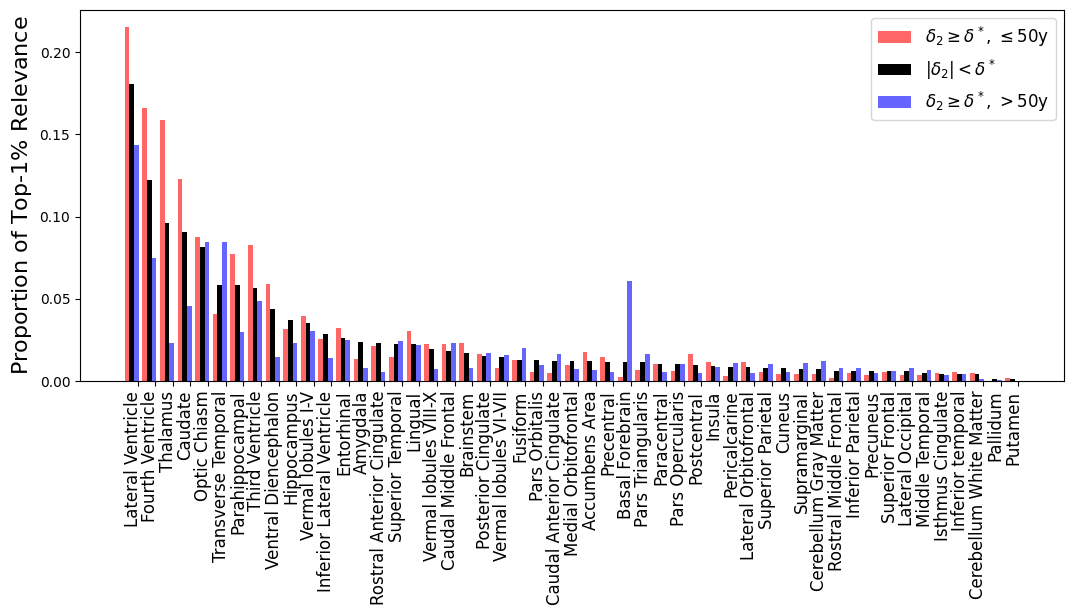}
            \caption{Distributions of Top-$1\%$ Relevance, via LRP$_2$, in young ($\leq50$y) individuals with high BAG ($\delta_2>\delta^*$), elderly individuals ($>50$y) with high BAG, and individuals with small-to-moderate BAG ($|\delta_2|<\delta^*$). Regions are ordered by descending proportion of T1R in the small-to-moderate BAG group.}
            \label{fig:LRP_2_dba}
        \end{figure}
        
        \begin{figure}[H]
            \centering
            \includegraphics[width=0.95\linewidth]{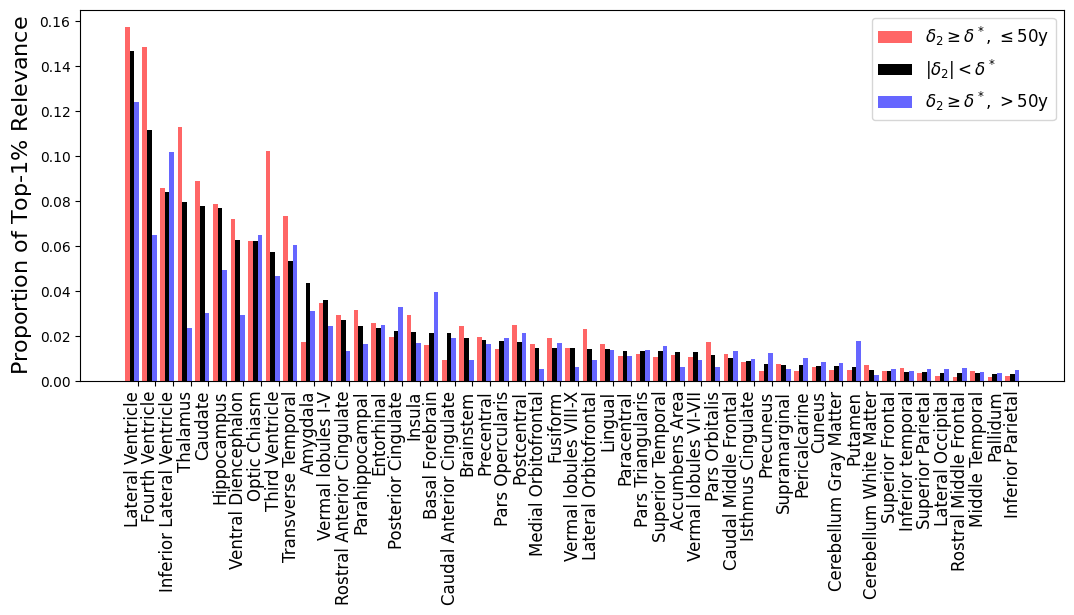}
            \caption{Distributions of Top-$1\%$ Relevance, via LRP$_3$, in young ($\leq50$y) individuals with high BAG ($\delta_2>\delta^*$), elderly individuals ($>50$y) with high BAG, and individuals with small-to-moderate BAG ($|\delta_2|<\delta^*$). Regions are ordered by descending proportion of T1R in the small-to-moderate BAG group.}
            \label{fig:LRP_3_dba}
        \end{figure}
        
    \section{Age-group Relevance}\label{app:age_brackets}
        
        \setcounter{figure}{0}
        \setcounter{table}{0}
        
\begin{figure}[H]
            \centering
            
            \begin{subfigure}[b]{\textwidth}
                \centering
                \includegraphics[width=\linewidth]{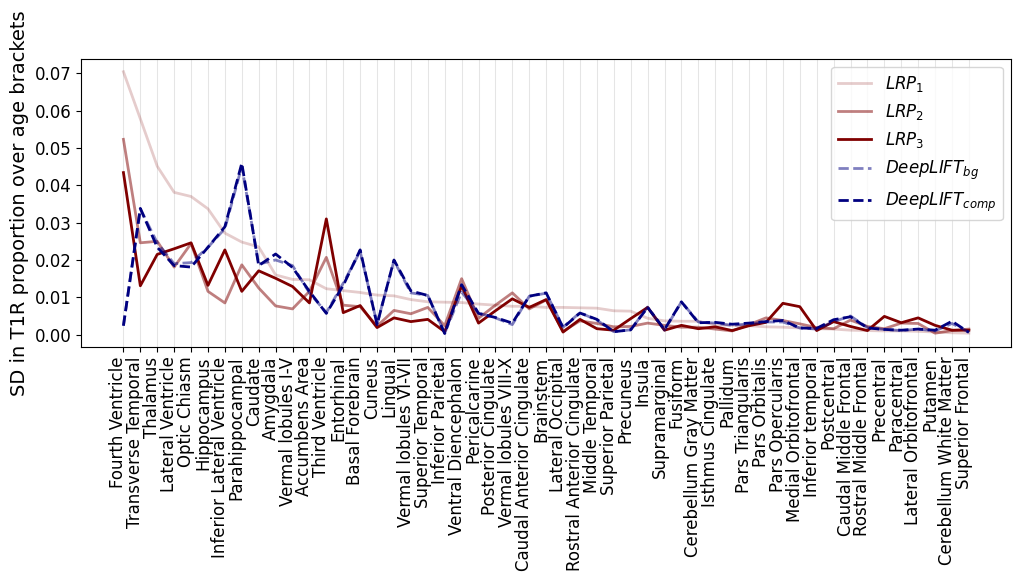}
                \caption{Standard Deviations in T1R over ages. Regions in descending order of SD for LRP$_1$.}
                \label{fig:SDs}
            \end{subfigure}
            ~
            \begin{subfigure}[b]{\textwidth}
                \centering
                \includegraphics[width=\linewidth]{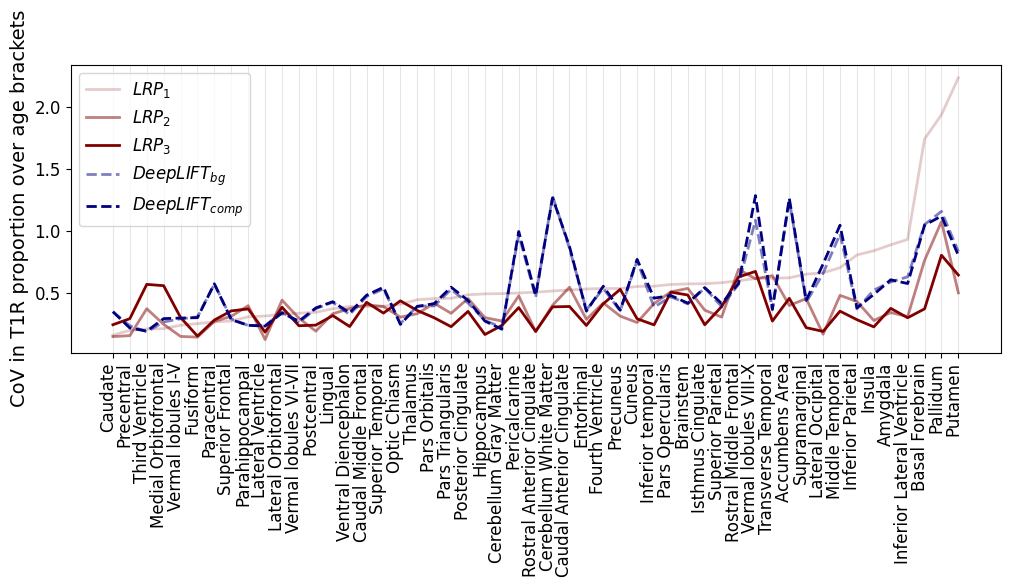}
                \caption{Coefficient of Variations in T1R over ages. Regions in ascending order of CoV for LRP$_1$.}
                \label{fig:CoVs}
            \end{subfigure}
            
            \caption{Changes in proportion of each region allocated T1R for each method, over age brackets}
            \label{fig:sd_and_cov}
        \end{figure}

        Figures (\ref{fig:SDs}) and (\ref{fig:CoVs}) show the Standard Deviations and Coefficients of Variation respectively of the proportion of T1R in each brain region across age brackets for each method.
        
        Regions with largest SD are those salient areas which change most in their assignment of relevance over ages. Regions with low SD often tend to be regions that are simply assigned very low proportions of T1R, such as the Cerebellum White Matter.
        
        We chose to use the CoV as a metric to find those salient regions which change least in their relevance assignment over ages. This is due to the fact that normalising the SD by the mean allows us to measure which regions have the least change in T1R relative to their mean T1R. The caudate nucleus, for example, has a large proportion of assigned T1R, and changes very little in this assignment across age brackets. On the other hand, regions with high CoV tend to be regions that are assigned low relevance overall. These regions can have small fluctuations in relevance assignment over age brackets, which leads to large CoVs. An example is the Putamen, which has very large CoV and very low proportions of assigned T1R (also, very small SD).
        
        \newpage
        
        \bibliography{ACAIN_brain_age_saliency_maps_paper.bib}

\begin{thebibliography}{10}
\providecommand{\url}[1]{\texttt{#1}}
\providecommand{\urlprefix}{URL }
\providecommand{\doi}[1]{https://doi.org/#1}

\bibitem{Ancona2018}
Ancona, M., Ceolini, E., {\"{O}}ztireli, C., Gross, M.: {Towards better
  understanding of gradient-based attribution methods for deep neural
  networks}. 6th International Conference on Learning Representations, ICLR
  2018 - Conference Track Proceedings pp. 1--16 (2018)

\bibitem{Anderton2002}
Anderton, B.H.: {Ageing of the brain}. Mechanisms of Ageing and Development
  (2002). \doi{10.1016/S0047-6374(01)00426-2}

\bibitem{Bach2015}
Bach, S., Binder, A., Montavon, G., Klauschen, F., M{\"{u}}ller, K.R., Samek,
  W.: {On pixel-wise explanations for non-linear classifier decisions by
  layer-wise relevance propagation}. PLoS ONE  \textbf{10}(7),  1--46 (2015).
  \doi{10.1371/journal.pone.0130140}

\bibitem{Cole2018}
Cole, J.H., Ritchie, S.J., Bastin, M.E., {Vald{\'{e}}s Hern{\'{a}}ndez}, M.C.,
  {Mu{\~{n}}oz Maniega}, S., Royle, N., Corley, J., Pattie, A., Harris, S.E.,
  Zhang, Q., Wray, N.R., Redmond, P., Marioni, R.E., Starr, J.M., Cox, S.R.,
  Wardlaw, J.M., Sharp, D.J., Deary, I.J.: {Brain age predicts mortality}.
  Molecular Psychiatry  (2018). \doi{10.1038/mp.2017.62}

\bibitem{Cole2017}
Cole, J.H., Poudel, R.P., Tsagkrasoulis, D., Caan, M.W., Steves, C., Spector,
  T.D., Montana, G.: {Predicting brain age with deep learning from raw imaging
  data results in a reliable and heritable biomarker}. NeuroImage
  \textbf{163},  115--124 (2017). \doi{10.1016/j.neuroimage.2017.07.059},
  \url{http://dx.doi.org/10.1016/j.neuroimage.2017.07.059}

\bibitem{Collins1999}
Collins, D., Zijdenbos, A., Baare, W., Evans, A.: {ANIMAL+INSECT: Improved
  Cortical Structure Segmentation}. IPMI Lecture Notes in Computer Science
  \textbf{1613/1999},  210--223 (1999).
  \doi{http://dx.doi.org/10.1007/3-540-48714-X_16}

\bibitem{Dinsdale2021}
Dinsdale, N.K., Bluemke, E., Smith, S.M., Arya, Z., Vidaurre, D., Jenkinson,
  M., Namburete, A.I.: {Learning patterns of the ageing brain in MRI using deep
  convolutional networks}. NeuroImage  \textbf{224},  117401 (jan 2021).
  \doi{10.1016/J.NEUROIMAGE.2020.117401}

\bibitem{Dombrowski2019}
Dombrowski, A.K., Alber, M., Anders, C.J., Ackermann, M., M{\"{u}}ller, K.R.,
  Kessel, P.: {Explanations can be manipulated and geometry is to blame}. In:
  Advances in Neural Information Processing Systems (2019)

\bibitem{Eitel2019}
Eitel, F., Soehler, E., Bellmann-Strobl, J., Brandt, A.U., Ruprecht, K., Giess,
  R.M., Kuchling, J., Asseyer, S., Weygandt, M., Haynes, J.D., Scheel, M.,
  Paul, F., Ritter, K.: {Uncovering convolutional neural network decisions for
  diagnosing multiple sclerosis on conventional MRI using layer-wise relevance
  propagation}. NeuroImage: Clinical  \textbf{24} (2019).
  \doi{10.1016/j.nicl.2019.102003}

\bibitem{Fonov2009}
Fonov, V., Evans, A., Almli, C., McKinstry, R., Collins, D.: {Unbiased
  nonlinear average age-appropriate brain templates from birth to adulthood}.
  NeuroImage  \textbf{47}, ~s102 (jul 2009).
  \doi{http://dx.doi.org/10.1016/S1053-8119(09)70884-5}

\bibitem{Geirhos2019}
Geirhos, R., Michaelis, C., Wichmann, F.A., Rubisch, P., Bethge, M., Brendel,
  W.: {Imagenet-trained CNNs are biased towards texture; increasing shape bias
  improves accuracy and robustness}. 7th International Conference on Learning
  Representations, ICLR 2019 (c),  1--22 (2019)

\bibitem{Grigorescu2019}
Grigorescu, I., Cordero-Grande, L., Edwards, A.D., Hajnal, J., Modat, M.,
  Deprez, M.: {Interpretable Convolutional Neural Networks for Preterm Birth
  Classification} pp.~1--4 (2019), \url{http://arxiv.org/abs/1910.00071}

\bibitem{Gunbey2014}
Gunbey, H.P., Ercan, K., Fjndjkoglu, A.S., Bulut, H.T., Karaoglanoglu, M.,
  Arslan, H.: {The Limbic Degradation of Aging Brain: A Quantitative Analysis
  with Diffusion Tensor Imaging}  (2014). \doi{10.1155/2014/196513},
  \url{http://dx.doi.org/10.1155/2014/196513}

\bibitem{Gupta2019}
Gupta, S., Chan, Y.H., Rajapakse, J.C.: {Decoding Brain Functional Connectivity
  Implicated in AD and MCI}. In: Lecture Notes in Computer Science (including
  subseries Lecture Notes in Artificial Intelligence and Lecture Notes in
  Bioinformatics) (2019). \doi{10.1007/978-3-030-32248-9_87}

\bibitem{He2016}
He, K., Zhang, X., Ren, S., Sun, J.: {Deep residual learning for image
  recognition}. In: Proceedings of the IEEE Computer Society Conference on
  Computer Vision and Pattern Recognition (2016). \doi{10.1109/CVPR.2016.90}

\bibitem{Hofmann2021}
Hofmann, S.M., Beyer, F., Lapuschkin, S., Loeffler, M., M{\"{u}}ller, K.R.,
  Villringer, A., Samek, W., Witte, A.V.: {Towards the Interpretability of Deep
  Learning Models for Human Neuroimaging}. bioRxiv p. 2021.06.25.449906 (jan
  2021). \doi{10.1101/2021.06.25.449906},
  \url{http://biorxiv.org/content/early/2021/08/26/2021.06.25.449906.abstract}

\bibitem{Jenkinson2002}
Jenkinson, M., Bannister, P., Brady, J., Smith, S.: {Improved optimisation for
  the robust and accurate linear registration and motion correction of brain
  images}. NeuroImage  \textbf{17}(2),  825--841 (2002)

\bibitem{Jenkinson2001}
Jenkinson, M., Smith, S.: {A global optimisation method for robust affine
  registration of brain images}. Medical Image Analysis  \textbf{5}(2),
  143--156 (2001)

\bibitem{Jonsson2019}
Jonsson, B.A., Bjornsdottir, G., Thorgeirsson, T.E., Ellingsen, L.M., Walters,
  G.B., Gudbjartsson, D.F., Stefansson, H., Stefansson, K., Ulfarsson, M.O.:
  {Brain age prediction using deep learning uncovers associated sequence
  variants}. Nature Communications  \textbf{10}(1) (2019).
  \doi{10.1038/s41467-019-13163-9},
  \url{https://doi.org/10.1038/s41467-019-13163-9}

\bibitem{Kingma2014}
Kingma, D.P., Ba, J.L.: {Adam: A Method for Stochastic Optimization}. 3rd
  International Conference on Learning Representations, ICLR 2015 - Conference
  Track Proceedings  (dec 2014), \url{https://arxiv.org/abs/1412.6980v9}

\bibitem{Kohlbrenner2019}
Kohlbrenner, M., Bauer, A., Nakajima, S., Binder, A., Samek, W., Lapuschkin,
  S.: {Towards best practice in explaining neural network decisions with LRP}
  pp.~1--5 (2019), \url{http://arxiv.org/abs/1910.09840}

\bibitem{Kossaifi2020}
Kossaifi, J., Kolbeinsson, A., Khanna, A., Furlanello, T., Anandkumar, A.:
  {Tensor Regression Networks}. Journal of Machine Learning Research
  \textbf{21},  1--21 (2020), \url{http://jmlr.org/papers/v21/18-503.html.}

\bibitem{Levakov2020}
Levakov, G., Rosenthal, G., Shelef, I., Raviv, T.R., Avidan, G.: {From a deep
  learning model back to the brain—Identifying regional predictors and their
  relation to aging}. Human Brain Mapping  \textbf{41}(12),  3235--3252 (aug
  2020). \doi{10.1002/HBM.25011/FORMAT/PDF}

\bibitem{Lutz2007}
Lutz, J., Hemminger, F., Stahl, R., Dietrich, O., Hempel, M., Reiser, M.,
  J{\"{a}}ger, L.: {Evidence of Subcortical and Cortical Aging of the Acoustic
  Pathway: A Diffusion Tensor Imaging (DTI) Study}. Academic Radiology
  \textbf{14}(6),  692--700 (jun 2007). \doi{10.1016/J.ACRA.2007.02.014}

\bibitem{Manera2020}
Manera, A., Dadar, M., Fonov, V., Collins, D.: {CerebrA, registration and
  manual label correction of Mindboggle-101 atlas for MNI-ICBM152 template}.
  Scientific Data (7(1)), ~1--9 (2020)

\bibitem{Montavon2017}
Montavon, G., Lapuschkin, S., Binder, A., Samek, W., M{\"{u}}ller, K.R.:
  {Explaining nonlinear classification decisions with deep Taylor
  decomposition}. Pattern Recognition  \textbf{65}(May 2016),  211--222 (2017).
  \doi{10.1016/j.patcog.2016.11.008},
  \url{http://dx.doi.org/10.1016/j.patcog.2016.11.008}

\bibitem{Peng2021}
Peng, H., Gong, W., Beckmann, C.F., Vedaldi, A., Smith, S.M.: {Accurate brain
  age prediction with lightweight deep neural networks}. Medical Image Analysis
   \textbf{68},  101871 (feb 2021). \doi{10.1016/J.MEDIA.2020.101871}

\bibitem{Peters2006}
Peters, R.: {Ageing and the brain} (2006). \doi{10.1136/pgmj.2005.036665}

\bibitem{Pianpanit2019}
Pianpanit, T., Lolak, S., Sawangjai, P., Ditthapron, A., Marukatat, S.,
  Chuangsuwanich, E., Wilaiprasitporn, T.: {Interpreting deep learning
  prediction of the Parkinson's disease diagnosis from SPECT imaging}  (aug
  2019), \url{https://arxiv.org/abs/1908.11199 http://arxiv.org/abs/1908.11199}

\bibitem{Raz2006}
Raz, N., Rodrigue, K.M.: {Differential aging of the brain: Patterns, cognitive
  correlates and modifiers} (2006). \doi{10.1016/j.neubiorev.2006.07.001}

\bibitem{Shafto2014}
Shafto, M.A., Tyler, L.K., Dixon, M., Taylor, J.R., Rowe, J.B., Cusack, R.,
  Calder, A.J., Marslen-Wilson, W.D., Duncan, J., Dalgleish, T., Henson, R.N.,
  Brayne, C., Bullmore, E., Campbell, K., Cheung, T., Davis, S., Geerligs, L.,
  Kievit, R., McCarrey, A., Price, D., Samu, D., Treder, M., Tsvetanov, K.,
  Williams, N., Bates, L., Emery, T., Erzin{\c{c}}lioglu, S., Gadie, A.,
  Gerbase, S., Georgieva, S., Hanley, C., Parkin, B., Troy, D., Allen, J.,
  Amery, G., Amunts, L., Barcroft, A., Castle, A., Dias, C., Dowrick, J., Fair,
  M., Fisher, H., Goulding, A., Grewal, A., Hale, G., Hilton, A., Johnson, F.,
  Johnston, P., Kavanagh-Williamson, T., Kwasniewska, M., McMinn, A., Norman,
  K., Penrose, J., Roby, F., Rowland, D., Sargeant, J., Squire, M., Stevens,
  B., Stoddart, A., Stone, C., Thompson, T., Yazlik, O., Barnes, D., Hillman,
  J., Mitchell, J., Villis, L., Matthews, F.E.: {The Cambridge Centre for
  Ageing and Neuroscience (Cam-CAN) study protocol: A cross-sectional,
  lifespan, multidisciplinary examination of healthy cognitive ageing}. BMC
  Neurology  (2014). \doi{10.1186/s12883-014-0204-1}

\bibitem{Shrikumar2017}
Shrikumar, A., Greenside, P., Kundaje, A.: {Learning important features through
  propagating activation differences}. 34th International Conference on Machine
  Learning, ICML 2017  \textbf{7},  4844--4866 (2017)

\bibitem{Shrikumar2016}
Shrikumar, A., Greenside, P., Shcherbina, A.Y., Kundaje, A.: {Not Just a Black
  Box : Learning Important Features Through Propagating Activation
  Differences}. In: Proceedings of the 33rd International Conference on
  MachineLearning (2016)

\bibitem{Sixt2019}
Sixt, L., Granz, M., Landgraf, T.: {When Explanations Lie: Why Many Modified BP
  Attributions Fail} (December 2019) (2019),
  \url{http://arxiv.org/abs/1912.09818}

\bibitem{Smilkov2017}
Smilkov, D., Thorat, N., Kim, B., Vi{\'{e}}gas, F., Wattenberg, M., Smilkov,
  D., Thorat, N., Kim, B., Vi{\'{e}}gas, F., Wattenberg, M.: {SmoothGrad:
  removing noise by adding noise}. arXiv p. arXiv:1706.03825 (jun 2017),
  \url{https://ui.adsabs.harvard.edu/abs/2017arXiv170603825S/abstract}

\bibitem{Smith2002}
Smith, S.: {Fast robust automated brain extraction.} Human Brain Mapping
  \textbf{17}(3),  143--155 (nov 2002)

\bibitem{Smith2004}
Smith, S., Jenkinson, M., Woolrich, M., Beckmann, C., Behrens, T.,
  Johansen-Berg, H., Bannister, P., De~Luca, M., Drobnjak, I., Flitney, D.,
  Niazy, R., Saunders, J., Vickers, J., Zhang, Y., De~Stefano, N., Brady, J.,
  Matthews, P.: {Advances in functional and structural MR image analysis and
  implementation as FSL}. NeuroImage  \textbf{23}(S1),  208--19 (2004)

\bibitem{Smith2019}
Smith, S.M., Vidaurre, D., Alfaro-Almagro, F., Nichols, T.E., Miller, K.L.:
  {Estimation of brain age delta from brain imaging}. NeuroImage  \textbf{200},
   528--539 (oct 2019). \doi{10.1016/J.NEUROIMAGE.2019.06.017}

\bibitem{Sowell2003}
Sowell, E.R., Peterson, B.S., Thompson, P.M., Welcome, S.E., Henkenius, A.L.,
  Toga, A.W.: {Mapping cortical change across the human life span}. Nature
  Neuroscience  (2003). \doi{10.1038/nn1008}

\bibitem{Sudlow2015}
Sudlow, C., Gallacher, J., Allen, N., Beral, V., Burton, P., Danesh, J.,
  Downey, P., Elliott, P., Green, J., Landray, M., Liu, B., Matthews, P., Ong,
  G., Pell, J., Silman, A., Young, A., Sprosen, T., Peakman, T., Collins, R.:
  {UK Biobank: An Open Access Resource for Identifying the Causes of a Wide
  Range of Complex Diseases of Middle and Old Age}. PLOS Medicine
  \textbf{12}(3),  e1001779 (mar 2015). \doi{10.1371/JOURNAL.PMED.1001779},
  \url{https://journals.plos.org/plosmedicine/article?id=10.1371/journal.pmed.1001779}

\bibitem{Sundararajan2016}
Sundararajan, M., Taly, A., Yan, Q.: {Gradients of Counterfactuals}  (nov
  2016), \url{https://arxiv.org/abs/1611.02639 http://arxiv.org/abs/1611.02639}

\bibitem{Sundararajan2017}
Sundararajan, M., Taly, A., Yan, Q.: {Axiomatic attribution for deep networks}.
  34th International Conference on Machine Learning, ICML 2017  \textbf{7},
  5109--5118 (2017)

\bibitem{Taylor2017}
Taylor, J.R., Williams, N., Cusack, R., Auer, T., Shafto, M.A., Dixon, M.,
  Tyler, L.K., Cam-CAN, Henson, R.N.: {The Cambridge Centre for Ageing and
  Neuroscience (Cam-CAN) data repository: Structural and functional MRI, MEG,
  and cognitive data from a cross-sectional adult lifespan sample}. NeuroImage
  \textbf{144},  262--269 (jan 2017). \doi{10.1016/J.NEUROIMAGE.2015.09.018}

\bibitem{Winkler2014}
Winkler, A., Ridgway, G., Webster, M., Smith, S., Nichols, T.: {Permutation
  inference for the general linear model}. NeuroImage  \textbf{92},  381--397
  (2014)

\bibitem{Zhao2019}
Zhao, L., Matloff, W., Ning, K., Kim, H., Dinov, I.D., Toga, A.W.: {Age-Related
  Differences in Brain Morphology and the Modifiers in Middle-Aged and Older
  Adults}  (2019). \doi{10.1093/cercor/bhy300}, \url{http://biobank.ctsu.ox.}

\end{thebibliography}
        \bibliographystyle{splncs04}

\end{document}